\documentclass[oldversion]{aa}
\usepackage{txfonts}
\usepackage{graphicx}
\usepackage{longtable}
\renewcommand\listofobjects{\relax}
\begin{document}
\title{Binary Stars in the Orion Nebula Cluster%
\thanks{Based on observations obtained at the European South\-ern
Observatory, La Silla, proposal number 68.C-0539, and at the W.M. Keck
Observatory, which is operated as a scientific partnership among the
California Institute of Technology, the University of California and
the National Aeronautics and Space Administration. The W.M.\ Keck
Observatory was made possible by the generous financial support of the
W.M.\ Keck Foundation.}}
\author{ Rainer K\"ohler\inst{1}
	\and
	 Monika G.\ Petr-Gotzens\inst{2}
	\and
	 Mark J.\ McCaughrean\inst{3,4}
	\and
	 Jerome Bouvier\inst{5}
	\and
	 Gaspard Duch\^ene\inst{5,6}
	\and
	 Andreas Quirrenbach\inst{7}
	\and
	 Hans Zinnecker\inst{4}
}
\offprints{Rainer K\"ohler, \email{koehler@strw.leidenuniv.nl}}

\institute{%
	Sterrewacht Leiden, P.O. Box 9513,
	NL-2300 RA Leiden, The Netherlands
\and
	European Southern Observatory,
	Karl-Schwarzschild-Str.\ 2, 85748 Garching bei M\"unchen, Germany
\and
	School of Physics, University of Exeter,
	Stocker Road, Exeter EX4 4QL, Devon, UK
\and
	Astrophysikalisches Institut Potsdam,
	An der Sternwarte 16, 14482 Potsdam, Germany
\and
	Laboratoire d'Astrophysique de Grenoble,
	Universit\'e Joseph Fourier, BP 53,
	38041 Grenoble Cedex 9, France
\and
	Department of Physics \& Astronomy,
	UCLA, Los Angeles, CA~90095-1562, USA
\and
	ZAH, Landessternwarte,
	K\"onigstuhl, 69117 Heidelberg, Germany
}
\date{Received 22 November 2005 / Accepted 26 July 2006}
\abstract{%
{}
{We report on a high-spatial-resolution survey for binary stars in the
  periphery of the Orion Nebula Cluster, at 5--15~arcmin (0.65 --
2\,pc) from the cluster center.}
{We observed 228 stars with adaptive optics systems, in order to find
  companions at separations of $0\farcs13$ -- $1\farcs12$ (60 --
  500\,AU), and detected 13 new binaries.  Combined with the results
  of Petr\ (1998), we have a sample of 275 objects, about half of
  which have masses from the literature and high probabilities
  to be cluster members.  We used an improved method to derive the
  completeness limits of the observations, which takes into account
  the elongated point spread function of stars at relatively large
  distances from the adaptive optics guide star.}
{The multiplicity of stars with masses $>2\,M_{\sun}$ is found to be
  significantly larger than that of low-mass stars.  The companion
  star frequency of low-mass stars is comparable to that of
  main-sequence M-dwarfs, less than half that of solar-type
  main-sequence stars, and 3.5 to 5 times lower than in the
  Taurus-Auriga and Scorpius-Centaurus star-forming regions.
  We find the binary frequency of low-mass stars in the periphery of
  the cluster to be the same or only slightly higher than for stars in
  the cluster core ($<$3~arcmin from $\theta^1$C~Ori).}
{This is in contrast to the prediction of the theory that the low
  binary frequency in the cluster is caused by the disruption of
  binaries due to dynamical interactions.  There are two ways out of
  this dilemma: Either the initial binary frequency in the Orion Nebula
  Cluster was lower than in Taurus-Auriga, or the Orion Nebula Cluster
  was originally much denser and dynamically more active.}
}
\keywords{stars: pre-main-sequence -- binaries: visual -- infrared:
stars -- surveys -- techniques: high angular resolution}
\maketitle

\section{Introduction}

Over the past decade it has become clear that stellar multiplicity can
be very high among young low-mass stars, with companion star
frequencies close to 100\,\% for young stars in well-known nearby
star-forming T~associations (Leinert et al.\ \cite{Leinert93}, Ghez et
al.\ \cite{Ghez93}, Ghez et al.\ \cite{Ghez97}, Duch\^ene
\cite{Duchene99a}).  Thus, our current understanding is that
star formation resulting in binary or multiple systems is very common,
if not the rule.  However, the multiplicity of low-mass
main-sequence field stars is significantly lower
, only $\sim 55\,\%$ for solar-type stars (Duquennoy
\& Mayor \cite{DM91}), and $\sim35$ to $42\,\%$ for M-dwarfs (Reid \&
Gizis \cite{RG97}, Fischer \& Marcy \cite{FM92}).

On the other hand, high binary frequencies are {\em not\/} observed
among low-mass stars in stellar clusters.  Binary surveys in the
center of the young Trapezium Cluster (e.g. Prosser et al.\
\cite{Prosser94}, Padgett et al.\ \cite{Padgett97}, Petr et al.\
\cite{Petr98}, Petr \cite{PetrPhD}, Simon et al.\ \cite{Simon99},
Scally et al.\ \cite{Scally99}, McCaughrean \cite{MJM2001}), which is
the core of the Orion Nebula Cluster (ONC); and in the young clusters
IC\,348 and NGC\,2024 (Duch\^ene et al.\ \cite{Duchene99b}, Beck et
al.\ \cite{BeckSC03}, Liu et al.\ \cite{Liu2003}, Luhmann et al.\
\cite{Luhman2005}), as well as those in older ZAMS clusters (Bouvier
et al.\ \cite{Bouvier97}, Patience et al.\ \cite{Patience98}) show
binary frequencies that are comparable to that of main-sequence fields
stars, i.e.\ lower by factors of 2 -- 3 than those found in loose T
associations.  The reason for this discrepancy is still
unclear. Theoretical explanations include:
\begin{itemize}
\item Disruption of cluster binaries through stellar encounters.

Kroupa (\cite{Kroupa95}) and Kroupa et al.\ (\cite{Kroupa99})
suggested that dynamical disruption of wide binaries (separations
${}>100\rm\,AU$) through close stellar encounters decreases the binary
fraction in clusters.  If the primordial binary output from the
star-formation process is the same in dense clusters and in loose
T~associations, then the number of ``surviving'' binaries depends on
the number of interactions of a binary system with other cluster
members that occurred since the formation of the cluster.  This number
is derived from the age of the cluster divided by the typical time between
stellar
interactions.  The typical time between interactions is inversely
proportional to the stellar volume density of the cluster (Scally et
al.\ \cite{Scally99}), thus binaries at the cluster center get
destroyed more quickly than those at larger radii.  Observing various
subregions of a single star-forming cluster representing different
stellar number densities will therefore reveal different binary
fractions if this mechanism is dominant in the evolution of binary
systems.

\item Environmental influence on the initial binary fraction.

Durisen \& Sterzik (\cite{DurSterz94}) and Sterzik et al.\
(\cite{Sterzik2003}) suggested an influence of the
molecular cores' temperature on the efficiency of the fragmentation
mechanism that leads to the formation of binaries.  Lower binary
fractions are predicted in warmer cores.  Assuming the ONC stars
formed from warmer cores than the members of the Taurus-Auriga
association, less initial binaries are produced in this scenario.
Observations of different subregions of the ONC should therefore
reveal the same (low) binary frequency (if the molecular cores in
these regions had the same temperature -- a reasonable, though
unverifiable assumption).
\end{itemize}
These theoretical concepts make different predictions that can be
tested observationally.  Indeed, measuring the binary fraction as a
function of distance to the cluster center will provide important
observational support for one or the other proposed theoretical
explanation.

\begin{table}[htp]
\caption[]{Fields observed for this work. Name is the designation in
Jones \& Walker (\cite{JW88}) or Parenago (\cite{Parenago}) of the
central star that was used to guide the adaptive optics system, $r$ is
the distance to $\theta^1$C~Ori.  In the last columns, we list the
number of stars actually observed in this field and the date(s) of
observation.}
\label{ObsTab}
\setlength{\tabcolsep}{3.5pt}
\begin{tabular}{lccrcc}
\noalign{\vskip1pt\hrule\vskip1pt}
Name   & $\alpha_{2000}$ & $\delta_{2000}$ & $r$ [\arcmin] & Targets & Obs.~Date\\
\hline
\object{JW0005} & 5:34:29.243	 & -5:24:00.37 	   & 11.8 &  6 & 09.12.2001 \\
\object{JW0014}	& 5:34:30.371	 & -5:27:30.46	   & 12.2 &  2 & 11.12.2001 \\
\object{JW0027}	& 5:34:34.012	 & -5:28:27.72	   & 11.8 &  4 & 11.12.2001 \\
\object{JW0045}	& 5:34:39.774	 & -5:24:28.27	   &  9.2 &  4 & 10.12.2001 \\
\object{JW0046}	& 5:34:39.917	 & -5:26:44.70	   &  9.7 &  5 & 10.12.2001 \\
\object{JW0050}	& 5:34:40.831	 & -5:22:45.07	   &  8.9 &  1 & 11.12.2001 \\
\object{JW0060}	& 5:34:42.187	 & -5:12:21.55	   & 14.0 &  2 & 10.12.2001 \\
\object{JW0064}	& 5:34:43.496	 & -5:18:30.01	   &  9.6 &  3 & 17.02.2002 \\
\object{JW0075}	& 5:34:45.188	 & -5:25:06.33 	   &  8.0 &  7 & 09.12.2001 \\
\object{JW0108}	& 5:34:49.867	 & -5:18:46.79	   &  8.1 &  3 & 10.12.2001 \\
\object{JW0116}	& 5:34:50.691	 & -5:24:03.25 	   &  6.5 &  5 & 11.12.2001 \\
\object{JW0129}	& 5:34:52.347	 & -5:33:10.38	   & 11.5 &  4 & 10.12.2001 \\
\object{JW0153}	& 5:34:55.390	 & -5:30:23.42	   &  8.8 &  2 & 17.02.2002 \\
\object{JW0157}	& 5:34:55.936	 & -5:23:14.58	   &  5.1 &  4 & 17.02.2002 \\
\object{JW0163}	& 5:34:56.560	 & -5:11:34.84	   & 12.8 &  3 & 11.12.2001 \\
\object{JW0165}	& 5:34:56.601	 & -5:31:37.48	   &  9.6 &  2 & 11.12.2001 \\
\object{JW0221}	& 5:35:02.202	 & -5:15:49.28	   &  8.4 &  4 & 11.12.2001 \\
\object{JW0232}	& 5:35:03.090	 & -5:30:02.27 	   &  7.5 &  5 & 09.12.2001 \\
\object{JW0260}	& 5:35:04.999	 & -5:14:51.45	   &  9.0 &  3 & 11.12.2001 \\
\object{JW0364}	& 5:35:11.455	 & -5:16:58.33	   &  6.5 & 13 & 09.12.2001 \\
\object{JW0421}	& 5:35:13.706	 & -5:30:57.76	   &  7.6 &  5 & 11.12.2001 \\
		&		 &		   &	  &    & 17.02.2002 \\
\object{JW0585}	& 5:35:18.275	 & -5:16:37.83	   &  6.8 & 17 & 09.12.2001 \\
\object{JW0666}	& 5:35:20.936	 & -5:09:15.92	   & 14.2 &  8 & 09.12.2001 \\
\object{JW0670}	& 5:35:21.043	 & -5:12:12.52	   & 11.2 &  2 & 17.02.2002 \\
\object{JW0747}	& 5:35:23.929	 & -5:30:46.82	   &  7.6 &  5 & 10.12.2001 \\
\object{JW0779}	& 5:35:25.417	 & -5:09:48.98	   & 13.8 & 11 & 09.12.2001 \\
\object{JW0790}	& 5:35:25.953	 & -5:08:39.42	   & 14.9 &  5 & 10.12.2001 \\
\object{JW0794}	& 5:35:26.121	 & -5:15:11.27	   &  8.5 &  6 & 09.12.2001 \\
\object{JW0803}	& 5:35:26.565	 & -5:11:06.84 	   & 12.5 &  4 & 10.12.2001 \\
\object{JW0804}	& 5:35:26.666	 & -5:13:13.97	   & 10.5 &  1 & 17.02.2002 \\
\object{JW0818}	& 5:35:27.716	 & -5:35:19.01	   & 12.3 &  2 & 11.12.2001 \\
\object{JW0866}	& 5:35:31.077	 & -5:15:32.23	   &  8.7 &  4 & 09.12.2001 \\
\object{JW0867}	& 5:35:31.116	 & -5:18:55.12	   &  5.8 &  3 & 09.12.2001 \\
\object{JW0873}	& 5:35:31.521	 & -5:33:07.91 	   & 10.5 &  6 & 11.12.2001 \\
\object{JW0876}	& 5:35:31.627	 & -5:09:26.88	   & 14.4 &  3 & 11.12.2001 \\
\object{JW0887}	& 5:35:32.487	 & -5:31:10.05	   &  8.8 &  1 & 17.02.2002 \\
\object{JW0915}	& 5:35:35.509	 & -5:12:19.38	   & 12.0 &  2 & 11.12.2001 \\
\object{JW0928}	& 5:35:37.385	 & -5:26:38.20 	   &  6.2 &  4 & 10.12.2001 \\
\object{JW0950}	& 5:35:40.519	 & -5:27:00.48 	   &  7.0 &  4 & 11.12.2001 \\
\object{JW0959}	& 5:35:42.019	 & -5:28:10.95	   &  8.0 &  5 & 10.12.2001 \\
\object{JW0963}	& 5:35:42.528	 & -5:20:11.73	   &  7.3 &  5 & 09.12.2001 \\
\object{JW0967}	& 5:35:42.803	 & -5:13:43.69	   & 11.7 &  1 & 17.02.2002 \\
\object{JW0971}	& 5:35:43.476	 & -5:36:26.22	   & 14.7 &  4 & 10.12.2001 \\
\object{JW0975}	& 5:35:44.558	 & -5:32:11.40	   & 11.3 &  3 & 11.12.2001 \\
\object{JW0992}	& 5:35:46.845	 & -5:17:54.69	   &  9.4 &  6 & 09.12.2001 \\
\object{JW0997}	& 5:35:47.326	 & -5:16:56.01	   & 10.1 &  4 & 11.12.2001 \\
\object{JW1015}	& 5:35:50.509	 & -5:28:32.56	   & 10.0 &  2 & 10.12.2001 \\
\object{JW1041}	& 5:35:57.831	 & -5:12:52.17	   & 14.8 &  1 & 17.02.2002 \\
\object{Par1605} & 5:34:47.201	 & -5:34:16.76	   & 13.1 &  3 & 10.12.2001 \\ 
\object{Par1744} & 5:35:04.822	 & -5:12:16.61	   & 11.5 &  4 & 11.12.2001 \\ 
\object{Par2074} & 5:35:31.223	 & -5:16:01.54 	   &  8.2 & 13 & 09.12.2001 \\ 
		 &		 &		   &	  &    & 17.02.2002 \\
\object{Par2284} & 5:35:57.539	 & -5:22:28.21	   & 10.3 &  2 & 17.02.2002 \\ 
\noalign{\vskip1pt\hrule}
\end{tabular}
\end{table}

\begin{figure*}[t]
\centerline{\includegraphics[angle=270,width=0.9\hsize]{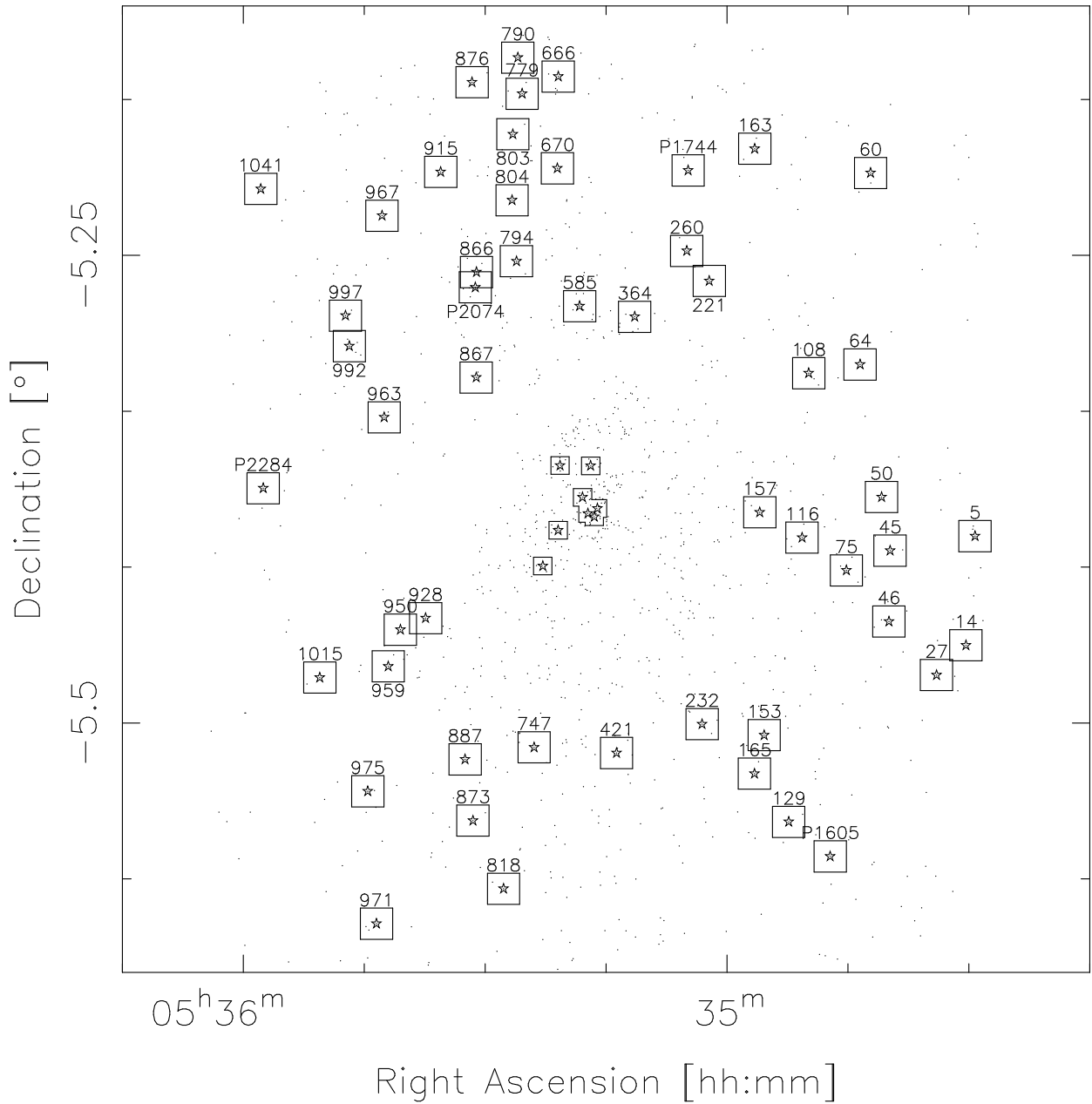}}
\caption{Fields observed in this work (large boxes) and Petr
(\cite{PetrPhD}, small boxes in the central region).  Dots mark the
positions of all stars in the list of Jones \& Walker (\cite{JW88}),
these are {\em not} the stars observed by us.
The stars used to guide the adaptive optics are marked by star symbols
and their number in Jones \& Walker or Parenago (\cite{Parenago}).
The large boxes show the areas where we selected our target stars.
These fields were not fully covered by our observations, we only
observed stars visible in the 2MASS quicklook images (see
Sect.~\ref{ObsSect}).}
\label{ObsFieldsFig}
\end{figure*}


The ONC is the best target for this study.  Its stellar population is
very well studied, more than 2000 members are known from extensive
near-infrared and optical imaging (Hillenbrand \cite{H97},
Hillenbrand \& Carpenter 2000).  To date, binary surveys of the
cluster have focused on the central 0.25\,pc core, where the stellar
density reaches as high as $2$ -- $5\times 10^4\rm\,pc^{-3}$
(McCaughrean \& Stauffer \cite{MJMStau94}, Hillenbrand \& Hartmann
\cite{Hillenbr98}).  The typical time between interactions for a
binary with a separation of 250\,AU ($\sim0\farcs5$ at the distance of
the ONC) in the core is $\sim 1\rm\,Myr$, the age of the cluster.
Therefore, most 250\,AU binaries are likely to have experienced at
least one close encounter.
However, for the observed stellar density distribution of the cluster,
which can be described by an isothermal sphere with $n \propto r^{-2}$
outside 0.06\,pc (Bate et al.\ \cite{Bate98}), the volume density at
$\sim$\,1\,pc distance from the center ($\sim8\arcmin$ at the distance
of the ONC) is roughly $200\rm\,pc^{-3}$ and the interaction
timescale for our 250\,AU binary would be $>250$\,Myr, hundreds of
times the age of the cluster.  We also know that dynamical mass
segregation in the cluster has not yet occurred (Bonnell \& Davies
\cite{BonnellDavies98}) and that the ejection of single stars from the
inner parts of the cluster (where many binary disruptions have already
occurred) has not been efficient to populate the outer regions if the
whole cluster is roughly virialized (Kroupa et al.\ \cite{Kroupa99}).
For these reasons, the binary fraction in
the outer parts of the ONC is unlikely to have been modified by the
dynamical evolution of the cluster, and should be the intrinsic value
resulting from the fragmentation process, while the binary frequency
of stars in the cluster core has already been lowered by dynamical
interactions.

We have measured the frequency of close binaries among stars in a
number of fields in the outer part of the ONC using adaptive optics
imaging in the K-band.  The results we obtain in this survey for the
outskirts of the Orion Nebula Cluster will be compared with a similar
study of the ONC core, carried out by Petr et al.\ (\cite{Petr98}) and
Petr (\cite{PetrPhD}). Since the same instrument and observing
strategy was used, we incorporate their results, and constrain the
radial distribution of the binary frequency from the ONC core to the
cluster's periphery.

\section{Observations}
\label{ObsSect}

{Since our study concentrates on the detection of close binary or multiple
stars with sub-arcsecond
separations, it is necessary to use an imaging technique that provides the
required performance in terms of high spatial resolution and sensitivity.
Therefore, we carried out adaptive optics near-infrared imaging
at ESO's 3.6\,m telescope and the Keck\,II telescope using their
respective adaptive optics instruments.

The choice of our target fields in the Orion Nebula cluster was
mainly guided by two criteria.
First, the necessity of having a nearby (within $\sim 30\arcsec$)
optically bright
reference star that controls the wavefront correction in order
to obtain (nearly) diffraction-limited images. Second, the fields should
be located at radii larger than 5arcmin from the cluster center.

{Based on these criteria, we searched the catalogues of
Jones \& Walker (\cite{JW88}, JW hereafter) and Parenago
(\cite{Parenago}) for stars with $\rm I<12.5^m$ and located
at radii of 5--15\,arcmin (0.65 -- 2\,pc) from the cluster center.
The resulting list contains 57 objects, 52 of them were observed in
the course of this project (see Tab.~\ref{ObsTab} and
Fig.~\ref{ObsFieldsFig}).  The remaining fields had to be omitted
because of time constraints.}}

Since the field of view's of the used instruments are at most
$12\farcs8\times12\farcs8$, it would have been very inefficient to
completely survey the fields around each guide star.  Instead, we used
quicklook-images from the 2MASS database to find stars within $30''$
from the guide stars, and pointed only at these sources.
Therefore, we did {\em not} observe all stars within $30\arcsec$ from
a guide star, but only those visible in 2MASS quicklook-images.
The 2MASS survey is complete to at least $\rm K_s = 14^{\rm mag}$
(except in crowded regions), so we can be sure all stars down to this
magnitude are visible in the quicklook-images and have been observed
by us.
Our estimated masses (Sect.~\ref{masssect}) indicate that this
probably already includes a few brown dwarfs, so the number of
interesting targets (low-mass stars) missed by this procedure is
negligible.

\begin{figure}[ht]
\includegraphics[width=0.49\hsize]{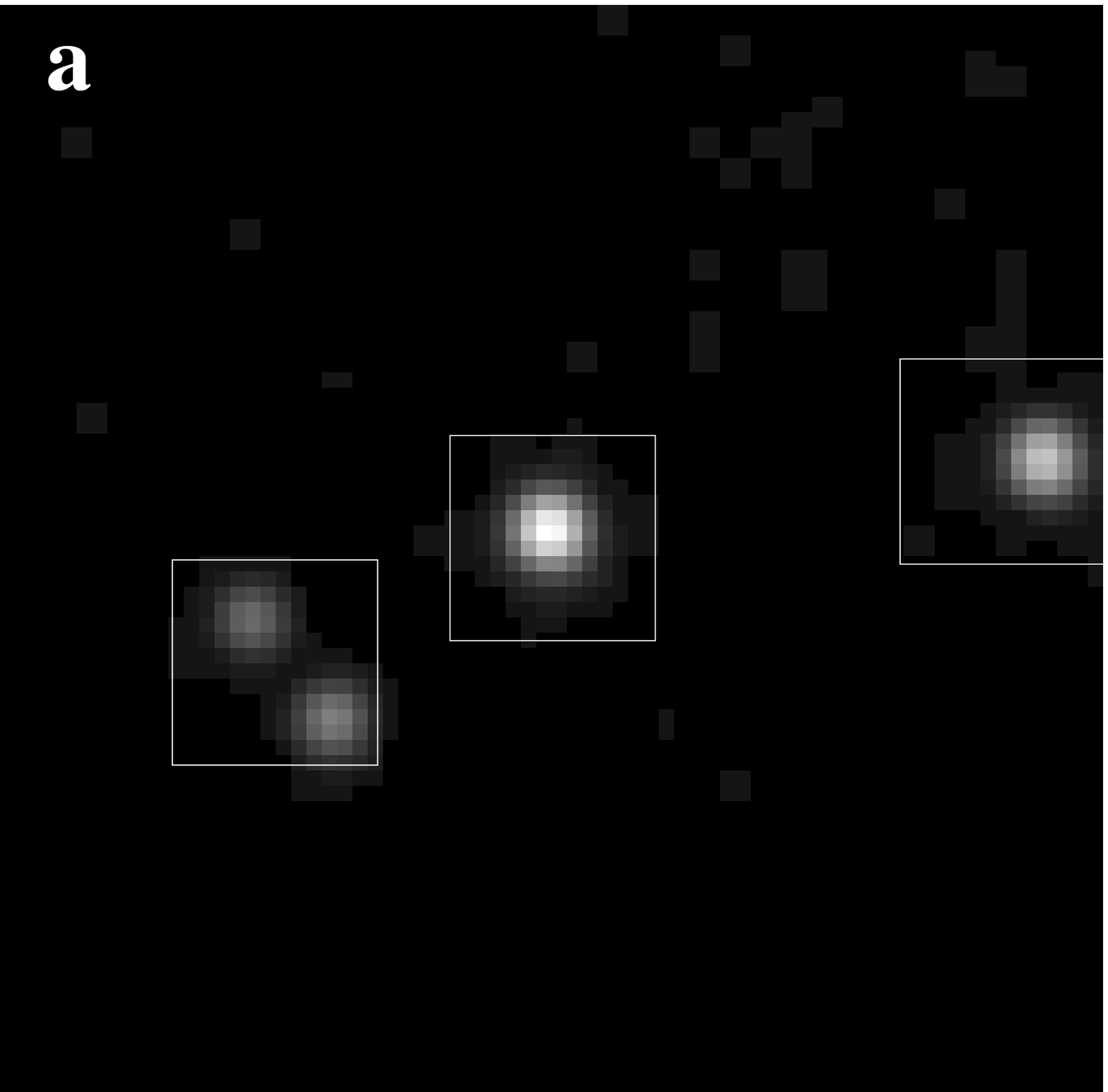}\hfill
\includegraphics[width=0.49\hsize]{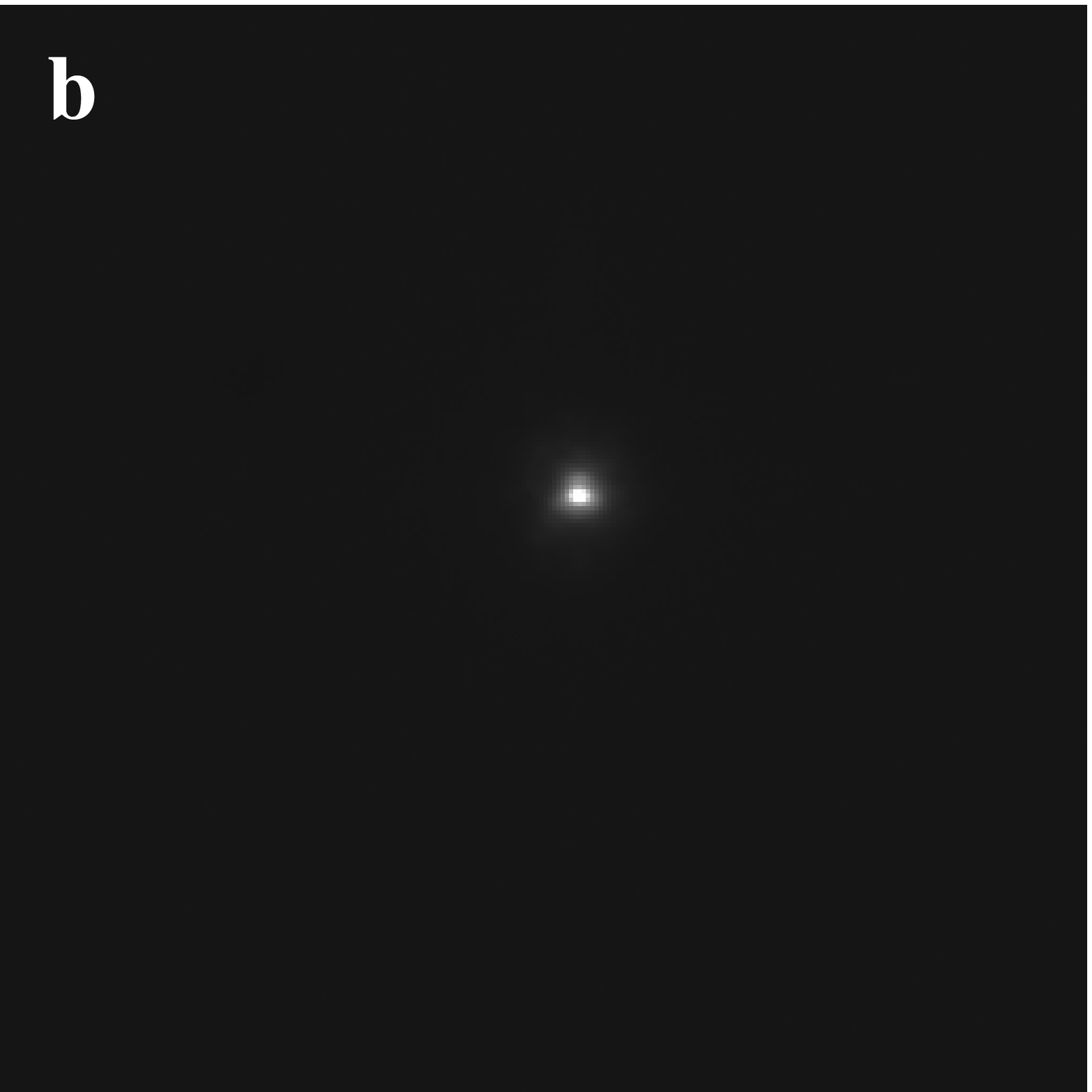}

\smallskip

\includegraphics[width=0.49\hsize]{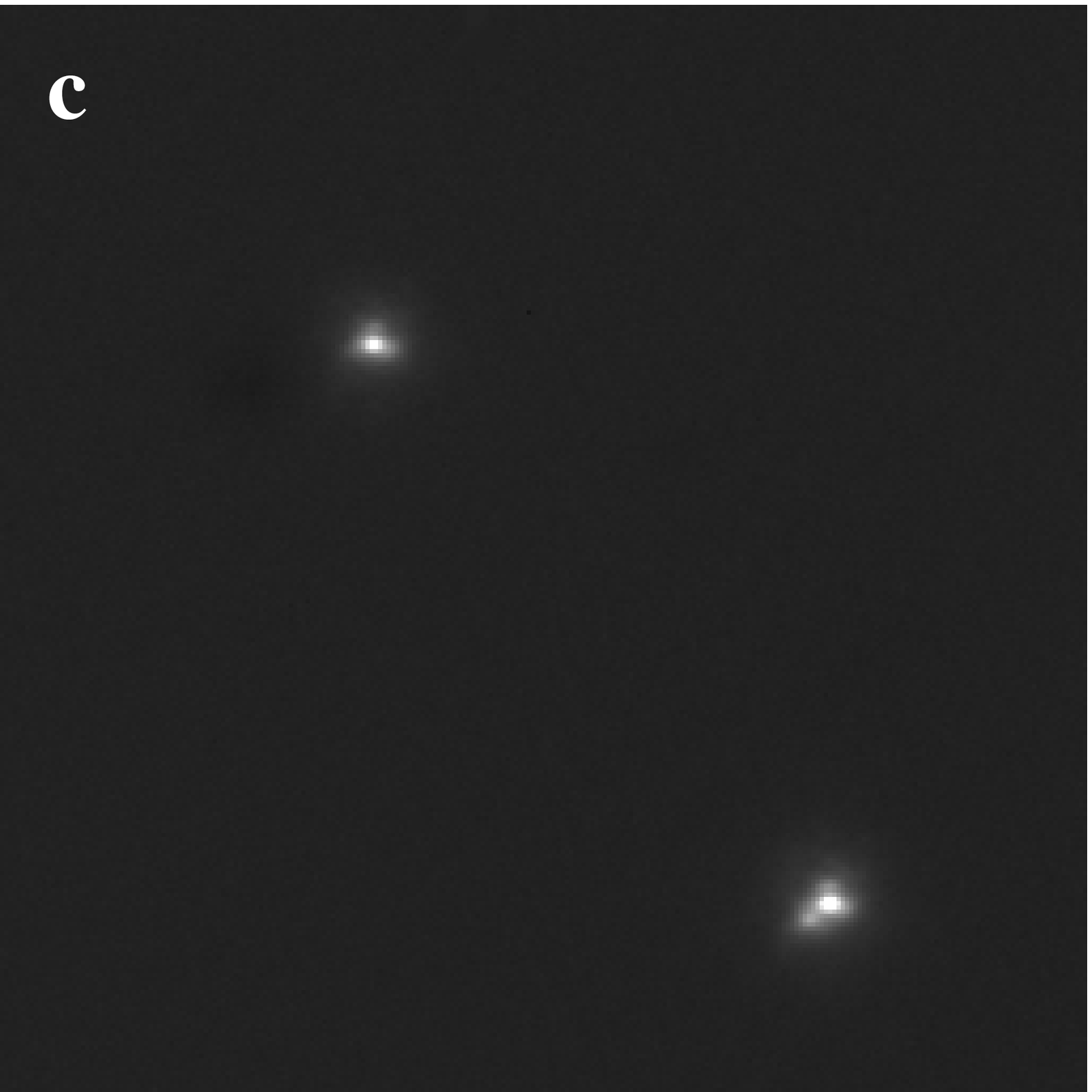}\hfill
\includegraphics[width=0.49\hsize]{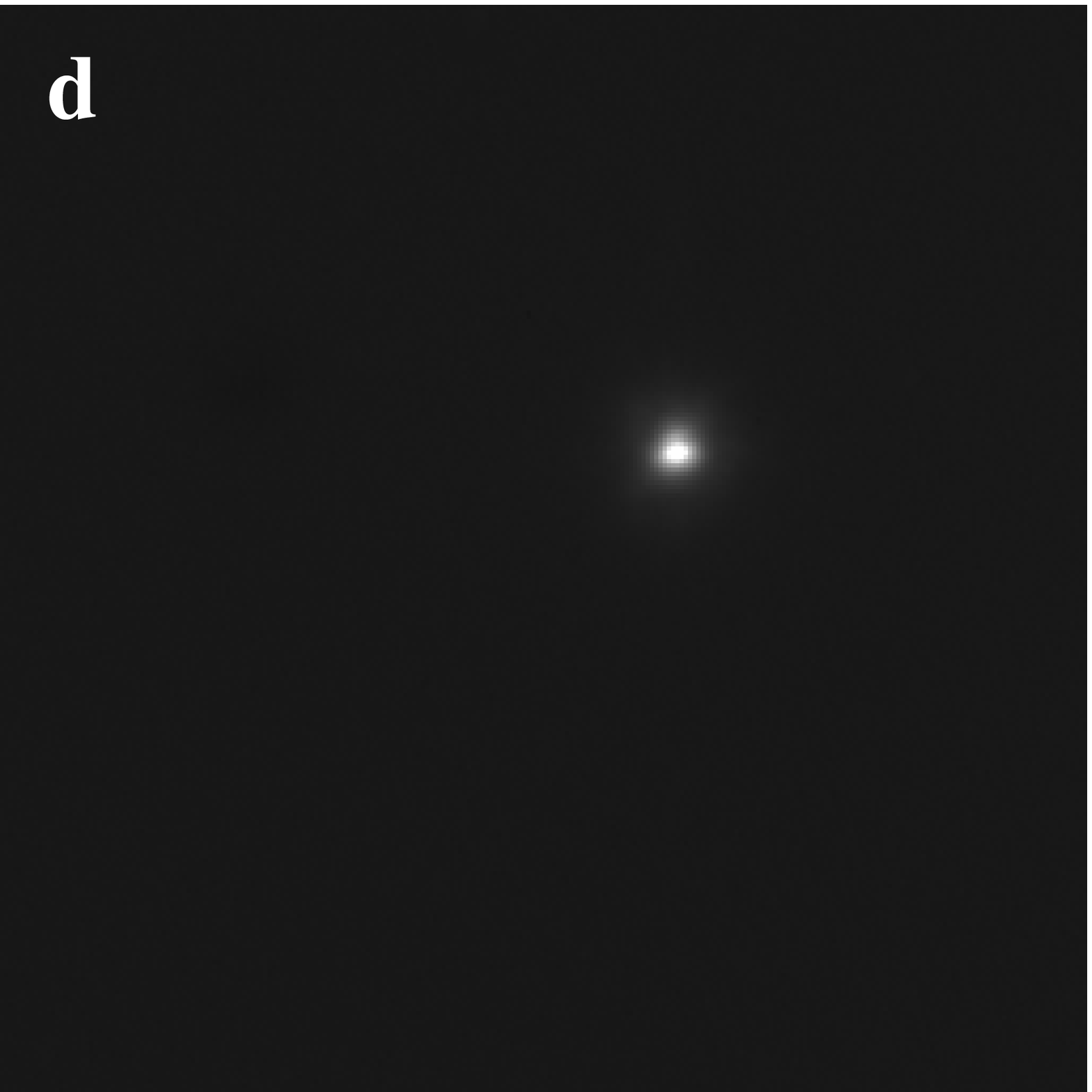}
\caption{The field around \object{JW0971} as an example of our
  observations. North is up and East is to the left in all images.
  The field-of-view in image a is about $70\times70\arcsec$, while the
  field-of-view in images b, c, and d is about $12\times12\arcsec$.
  a: The 2MASS-quicklook image centered on JW0971.  The positions of
     the images obtained with the AO system ADONIS are outlined.
  b: The AO-corrected image centered on the guide star, JW0971.
  c: The AO-corrected image of the two stars south-east of JW0971.
     The star to the southwest is \object{JW0974} and has a companion
     at $0.32\arcsec$ separation (cf.~Tab.~\ref{BinTab}).
  d: The AO-corrected image centered on the star west of JW0971.}
\label{ExamplObsFig}
\end{figure}

Figure~\ref{ExamplObsFig} shows as an example the 2MASS
quicklook-image of the region around JW0971, and our
AO-corrected images.


\subsection{ESO\,3.6m/ADONIS observations}
The major part of the observations was carried out during three nights
in December 2001 at the ESO 3.6\,m telescope on La Silla.  We used the
adaptive optics system ADONIS with the SHARP\,II camera to obtain
nearly-diffraction limited images in the K$_s$-band (2.154$\rm\mu m$).
The pixel scale was 50\,mas/pixel, which provides full Nyquist
sampling at 2.2$\rm\mu m$.  Since SHARP\,II is equipped with a
256$\times$256 pixels NICMOS\,3 array detector, the field-of-view is
$12\farcs8\times12\farcs8$.  During the first observing night, we
typically recorded for each target star 20 frames with 3\,sec
individual exposure time, while during the second and third night 20
frames of 5\,sec each were taken per object.  After all targets near
one guide star were observed (which took on average 4, at most 9
pointings), we observed an empty field at about one arcminute distance
from the guide star to determine the sky background.  Measuring the
sky background every 20\,minutes is enough for the purpose of this
project, i.e.\ astrometry.\footnote{The photometric measurements
obtained in the course of the data reduction were used only to solve
ambiguous cases in the identification of our objects in photometric
catalogues.}

\subsection{KeckII/NIRC2 observations}
In February 2002, eleven fields were observed in one half night at the
Keck\,II telescope.  We used the adaptive optics system and the
NIRC\,2 camera with the K$_p$ filter (2.124$\rm\mu m$).  NIRC\,2
offers the choice of three different pixels scales: 10, 20, or
40\,mas/pixel.  The 40\,mas/pixel scale does not allow Nyquist
sampling in the K-band, and the 20\,mas/pixel optics has significant
field curvature, therefore we decided to use the 10\,mas/pixel camera
in order to achieve the full resolution over the whole field-of-view.
The detector has a 1024$\times$1024 Aladdin-3 array, so the
field-of-view is about 10\arcsec$\times$10\arcsec.  The integration
time was 60\,sec per target (20 co-adds of 3\,sec each).  The one
exception is Parenago~2074, which was saturated in 3\,sec.  Here we
took more frames with 0.25\,sec exposure time, resulting in about
20\,sec total integration time.  Exposures of the sky background
were done in the same way as during the observations with ADONIS.

\begin{table*}[t]
\caption[]{New ONC Binaries found in this work. Names with ``JW'' are
from Jones \& Walker \cite{JW88}, those with ``AD95'' from Ali \&
Depoy \cite{AD95}, and Par2074 is from Parenago (\cite{Parenago}).
The distance to $\theta^1$C is listed as $r$ like in Tab.~\ref{ObsTab}.
The last three columns list identification, mass from H97 and from
IR~photometry, and membership probability given in H97.}
\label{BinTab}
\begin{tabular}{llcrccrccc}
\noalign{\vskip1pt\hrule\vskip1pt}
Name	&\hfil$\alpha_{2000}$ & $\delta_{2000}$ & $r$ [$'$]
							& Separation [$''$] & $\Delta K$ [mag] & H97 id & H97 mass& IR mass& Mem.\ [\%]\\
\hline
\object{JW0235}	   & 5:35:03.586 & -5:29:27.06 &  6.87	& $0.48\pm0.01$ &	$0.47\pm0.15$ &	 235~	&	& 0.5 &	   \\ 
\object{JW0260}	   & 5:35:04.999 & -5:14:51.45 &  9.00	& $0.33\pm0.01$ &	$0.17\pm0.05$ &	 260~	& 4.13	& 3.5 & 97 \\ 
\object{AD95 2380} & 5:35:12.328 & -5:16:34.04 &  6.89	& $0.59\pm0.02$ &	$2.99\pm0.15$ &		&	& 3.0 &    \\ 
\object{JW0406}	   & 5:35:13.299 & -5:17:09.98 &  6.27	& $0.94\pm0.01$ &	$1.19\pm0.06$ &	 406~	& 0.21	& 1.5 & 99 \\ 
\object{JW0566}	   & 5:35:17.756 & -5:16:14.68 &  7.15	& $0.85\pm0.02$ &	$1.73\pm0.06$ &	 566~	&	&     &  0 \\ 
\object{AD95 1468} & 5:35:17.352 & -5:16:13.62 &  7.16	& $1.08\pm0.01$ &	$0.10\pm0.05$ &		&	&     &    \\ 
\object{JW0765}	   & 5:35:24.892 & -5:09:27.87 & 14.08	& $0.35\pm0.02$ &	$0.06\pm0.10$ &	 765~	& 0.19	& 0.5 & 97 \\ 
\object{JW0767}	   & 5:35:25.087 & -5:15:35.73 &  8.08	& $1.10\pm0.01$ &	$1.99\pm0.10$ &	 767~	& 0.26	& 1.0 & 99 \\ 
\object{JW0804}	   & 5:35:26.666 & -5:13:13.97 & 10.47	& $0.40\pm0.01$ &
							     $4.1\phantom{0}\pm0.2\phantom{0}$ & 804~	& 1.65	& 1.2 &  0 \\ 
\object{JW0876}	   & 5:35:31.627 & -5:09:26.88 & 14.44	& $0.49\pm0.01$ &	$0.50\pm0.05$ &	 876~	& 0.84	& 3.5 & 13 \\ 
\object{JW0959}	   & 5:35:42.019 & -5:28:10.95 &  7.99	& $0.34\pm0.01$ &	$0.07\pm0.03$ &	 959~	& 2.41	& 2.7 & 97 \\ 
\object{JW0974}	   & 5:35:44.417 & -5:36:38.02 & 14.98	& $0.32\pm0.02$ &	$1.41\pm0.05$ &	 974~	& 0.14	& 0.3 & 99 \\ 
\object{Par2074}   & 5:35:31.223 & -5:16:01.54 &  8.23	& $0.47\pm0.01$ &	$3.23\pm0.10$ & 2074~	& 16.3\quad\null & $>$4 & 99 \\ 
\noalign{\vskip1pt\hrule}
\end{tabular}
\end{table*}


\begin{figure*}[ht]
\includegraphics[angle=270,width=\hsize]{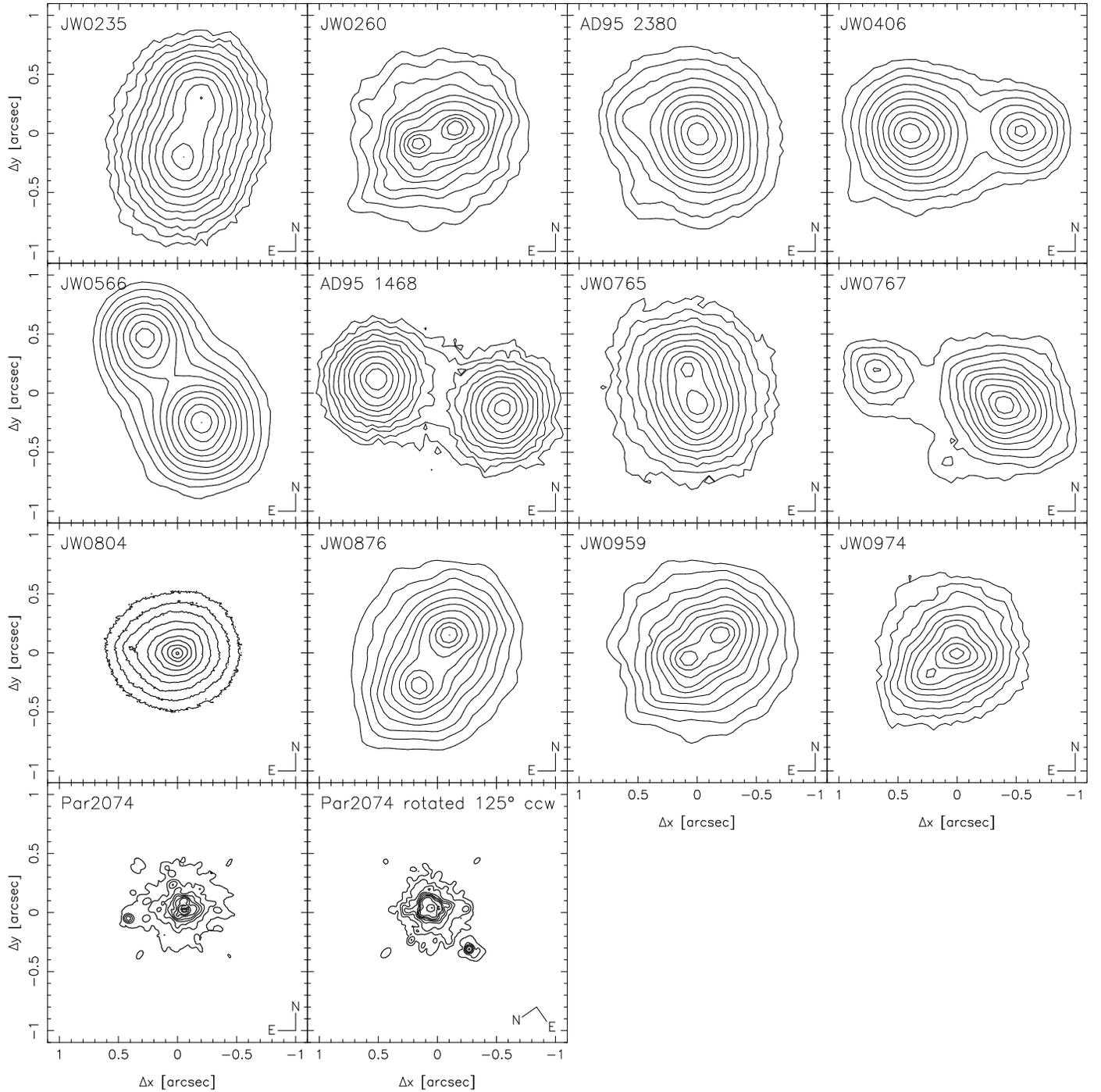}
\caption{Images of the 13 binaries found in this work, shown with a
logarithmic scale. North is up and east is left, except for the last
image.  Because Parenago 2074 is the brightest star we observed with
the Keck telescope, it suffers from unusually strong diffraction
spikes, and it was unclear from a single image whether the candidate
companion we had found was a physical object or an instrumental
artefact. We therefore rotated the field-of-view by 125$^\circ$ and
repeated the observation. The companion rotated by the same angle,
which proves its physical nature.}
\label{BinContourFig}
\end{figure*}


\section{Data Reduction}
Images were sky-subtracted with a median sky image and flat fielded by
an appropriate flat defined by dome-flat illumination.
Bad pixels were replaced by the median of the closest good neighbors.
Finally images were visually inspected for any artifacts or residuals.

The daophot package within IRAF\footnote{IRAF is distributed by the
    National Optical Astronomy Observatories, which are operated by
    the Association of Universities for Research in Astronomy, Inc.,
    under cooperative agreement with the National Science Foundation.}
was used to identify stars and measure their positions and magnitudes.
First, we used the daofind task to identify stars.  All images were
inspected visually to confirm the detections, the parameters of
daofind (in particular, the threshold) were adjusted to make sure no
stars were missed by the automatic procedure.  Then we used the phot
task to carry out aperture photometry of the sources found.  The
aperture radius was 1\arcsec\ in order to make sure no flux was lost
in the wings of the PSF.

We count pairs of stars separated by $1\farcs12$ (500\,AU) or less as
binary candidates (not all of them are gravitationally bound
  binaries, see Sect.~\ref{BgSect}).  Daofind is unable to reliably
  detect binary sources with such small separations, therefore all
  detected stars had to be inspected visually to identify binaries.
  An aperture with 1\arcsec\ radius measures flux from both
  components in these cases, therefore the phot task was run again on
  the binaries, with apertures adjusted for each individual binary to
  measure the individual fluxes without contamination of the secondary
  by the primary and vice versa.
  The results of these measurements were used to compute the exact
  separations, position angles, total fluxes, and flux ratios of the
  binary stars.


\section{Results}

\subsection{Stellar detections and Binaries}
In total, we observed 228 stars in 52 fields (see Tab.~\ref{ObsTab}
and Fig.~\ref{ObsFieldsFig})\footnote{The complete target list
is available online.}.  Stars located within $1\arcsec$ from
one of the edges of the images are not counted, to ensure that our
binary census is as unbiased as possible.

We are sensitive to companions at separations in the range
$0\farcs13$ -- $1\farcs12$ or 60 -- 500\,AU at the distance of the ONC
(450\,pc).  The lower limit is the diffraction limit of the 3.6\,m
telescope in K, the outer limit was chosen to limit the number of
chance alignments of unrelated stars (see section~\ref{BgSect}).
These limits are the same as those used by Petr (\cite{PetrPhD}) for
their binary survey in the center of the ONC.

We find 13 binary candidates in the separation range $0\farcs13$ --
$1\farcs12$.  No higher-order multiples were found, which is not
surprising given the limited separation range.  Table~\ref{BinTab}
lists the binaries, their separations, magnitude differences, their ID
number in Hillenbrand (\cite{H97}), mass estimates and probability to
be cluster members.  Figure~\ref{BinContourFig} shows images of the
binary stars.

Some of the binaries have a rather low membership probability.  On the
one hand, it would not be surprising to find binary foreground stars.
On the other hand, there is some evidence that membership
probabilities based on proper motions may be underestimated.
Hillenbrand~(\cite{H97}) report that several of the sources identified
as being externally ionized by the massive Trapezium stars (and,
therefore, physically close to these stars) are designated as proper
motion non-members.  Furthermore, the proper motions of undetected
binaries can have systematic errors due to orbital motion or
photometric variability (Wielen \cite{Wielen97}), leading to
misclassification as non-members.
In the analysis in Sect.~\ref{DiscuSect}, we will consider two
samples: one containing only stars with membership probabilities
higher than 50\,\%, and one with all stars for which mass estimates
could be obtained, including those with zero membership probability.


\begin{figure*}[t]
\centerline{\includegraphics[angle=270,width=\hsize]{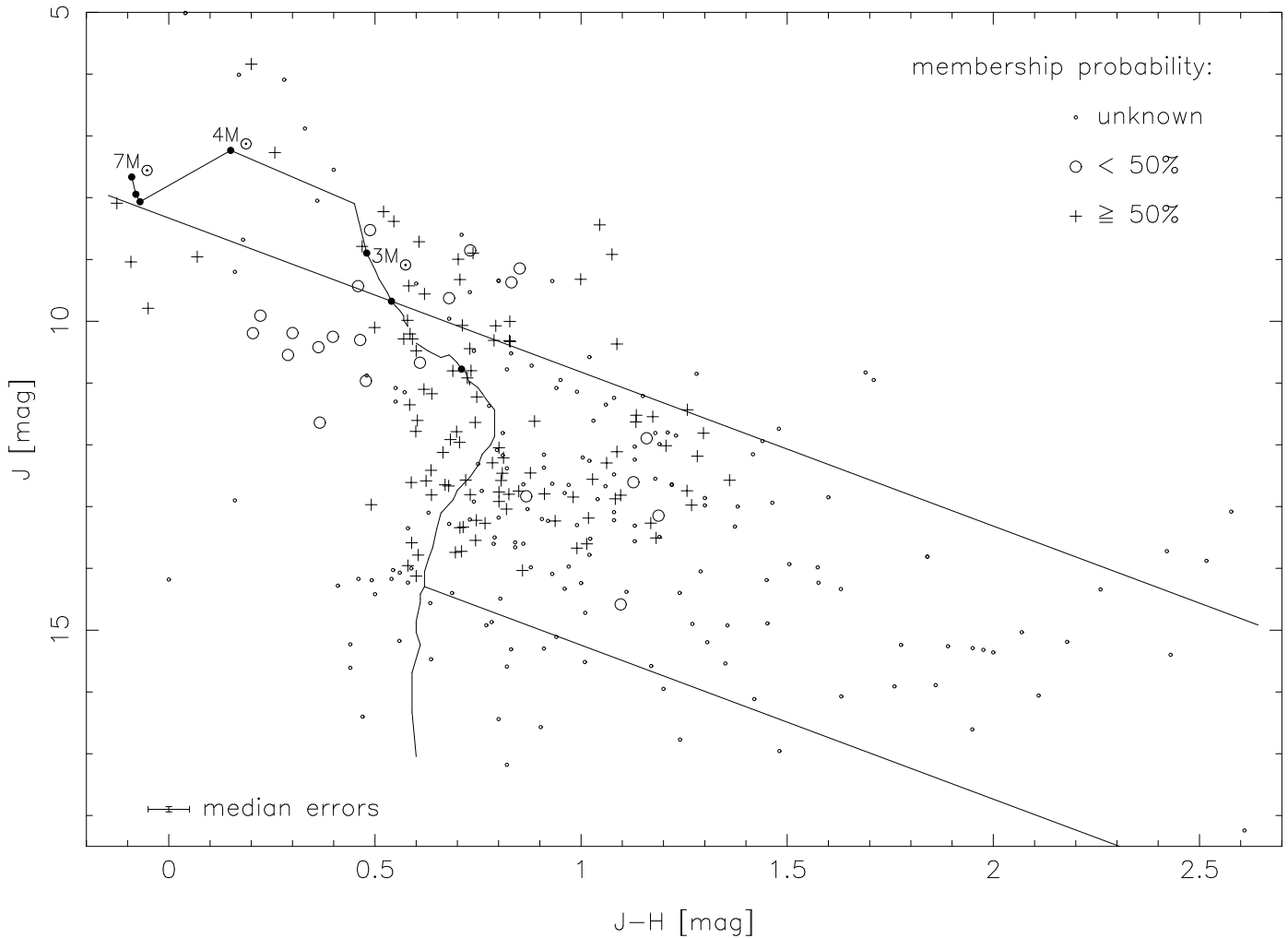}}
\caption{Color-Magnitude diagram for ONC stars with IR-photometry from
2MASS or Muench et al.\ (\cite{Muench2002}).  The symbols denote
proper motion members, non-members, and stars without membership
information.  Median errors are indicated in the lower left.
Also shown is the theoretical 1\,Myr isochrone of the
models of Baraffe et al.\ (\cite{Baraffe98}) and Siess et al.\
(\cite{Siess2000}).  The transition from one set of models to the
other between 1.4$\,M_{\sun}$ and 1.5$\,M_{\sun}$ is indicated by a
gap in the isochrone line between $\rm J=10.1^m$ and
$\rm 10.4^m$.  The dots on the isochrone mark the 1, 2, 3 \dots
7$\,M_{\sun}$ models.  The straight lines show the extinction vectors
through the 0.1 and 2$\,M_{\sun}$ models.
The isochrone turns downward again for masses${}>4\,M_{\sun}$, making
it impossible to distinguish a highly-extincted star with, e.g.,
4.5$\,M_{\sun}$ from a less-extincted star of $\sim3\,M_{\sun}$.
The extinction vector through the 2$\,M_{\sun}$ model has been
extended backwards to demonstrate that this ambiguity does not exist
for masses${}<2\,M_{\sun}$.  We adopt $2\,M_{\sun}$ as limit to classify
our stars as low-mass or intermediate- to high-mass.}
\label{ColMagJHFig}
\end{figure*}

\subsection{Selection of the Low- and Intermediate- to High-mass Samples}
\label{masssect}

The multiplicity of high- and low-mass stars in the ONC is
significantly different (Sect.~\ref{MassSect}, Preibisch et al.\
\cite{Preibisch99}, Preibisch et al.\ \cite{Preibisch2001}, Schertl et
al.\ \cite{Schertl03} and references therein).  Therefore,
we have to select suitable subsamples of our
target list in order to obtain meaningful results.  Probably the best
mass estimates for stars in the ONC were given by Hillenbrand
(\cite{H97}, H97 hereafter), who used spectroscopic and photometric
data to create an HR diagram.  Comparison with theoretical
pre-main-sequence evolutionary tracks of D'Antona \& Mazzitelli
(\cite{DanMaz94}) yielded the mass and age of each star.  From their
work, we can get masses for 126 stars in our sample, and 40 stars in
the sample of Petr (\cite{PetrPhD}).
The accuracy of masses predicted by evolutionary tracks is limited, as
most models are only marginally consistent with masses determined from
measured orbital dynamics (Hillenbrand \& White \cite{HillWhite2004}).
However, in this work the mass estimates are only used to distinguish
between intermediate- to high-mass stars, low-mass stars, and
potential sub-stellar objects, so the results are not affected
by systematic discrepancies.

To obtain mass estimates for a larger fraction of our sample, we use
2MASS photometry in the J- and H-band to construct a color-magnitude
diagram (CMD, Fig.~\ref{ColMagJHFig}).  We decided to use J and H
because the photometry in these bands is less affected by infrared
excess emission of circumstellar material than K-band photometry.  To
estimate masses, we de-redden the stars along the standard extinction
vector for Orion (Johnson \cite{Johnson67}) and search the
intersection of the vector with the 1\,Myr isochrone of the
theoretical pre-main-sequence tracks of Baraffe et al.\
(\cite{Baraffe98}) for masses below 1.4$\,M_{\sun}$ and Siess et al.\
(\cite{Siess2000}) for masses above 1.4$\,M_{\sun}$.  The agreement
between these two sets of tracks is reasonably good, and the
transition point has been chosen to minimise the difference.

In cases where a star falls to the left (blueward) of the isochrone,
we adopt the mass of the model that has the same J-magnitude, i.e.\ we
do not follow the direction of the extinction vector, but shift the
star horizontally onto the isochrone.  This procedure is reasonable,
since the measurement error of $J-H$ is much larger than that of $J$
(see the error bars in the lower left of
 Fig.~\ref{ColMagJHFig})\footnote{Many of the stars to the left of the
isochrone have a low probability to be members of the Orion Nebula
Cluster (cf.\ JW)}. 
In this way, we can obtain mass estimates for 173 stars.  Combined
with the results from H97, we have mass estimates for 187 stars.

For the selection of the sub-samples, we used the following limits:
Stars with masses of at least 2\,$M_{\sun}$ are the intermediate- to
high-mass sample (28 stars, 4 binaries), those with masses between 0.1
and 2\,$M_{\sun}$ form the low-mass sample (146 stars, 7 binaries),
and we designate objects with mass estimates below 0.1\,$M_{\sun}$ as
sub-stellar candidates (14 objects, no binaries).  Changing the upper
limit of the low-mass sample to, e.g., 1.3\,$M_{\sun}$ does not change
our conclusions, it only reduces their statistical significance.

A similar procedure was carried out for the target list of Petr
(\cite{PetrPhD}), using infrared photometry from Muench et al.\
(\cite{Muench2002}), which yielded 124 mass estimates: 22
intermediate- to high-mass stars (6 binaries), 83 low-mass stars (82
systems, 6 binaries), and 19 sub-stellar candidates (no binaries).

\begin{figure}[t]
\centerline{\includegraphics[angle=270,width=\hsize]{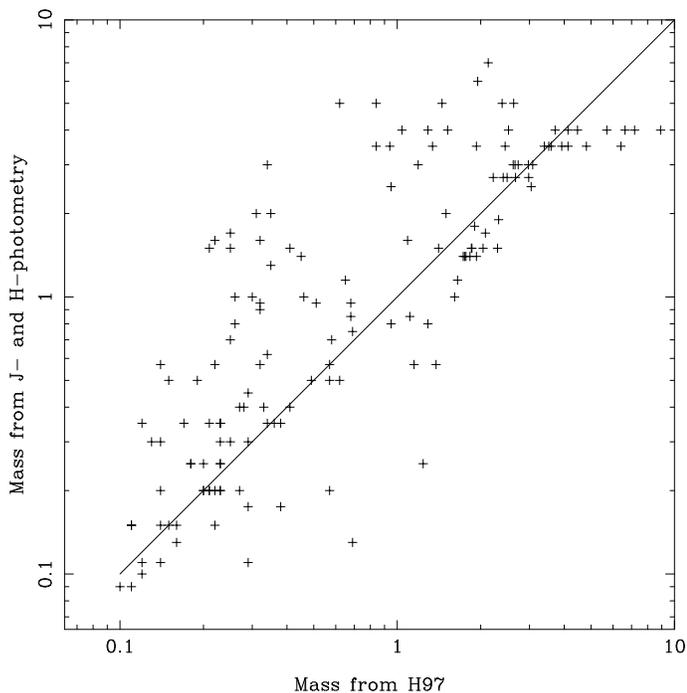}}
\caption{Comparison of masses from H97 and estimated from 2MASS or
  Muench et al.\ (\cite{Muench2002}) J- and H-photometry.  The line
  marks the points where both mass estimates are equal.}
\label{cmpMassJHFig}
\end{figure}

For 152 stars, we have masses from both H97 and IR-photometry, Fig.\
\ref{cmpMassJHFig} shows a comparison of mass estimates obtained in
both ways.  On average, the method using J and H photometry
overestimates the masses compared to H97 by a factor of 1.3.  This
could be caused by a number of reasons, for example infrared excess
due to circumstellar disks.  While the contribution of disks is
smaller for shorter wavelengths (which is the reason why we choose to
estimate masses from J- and H-band fluxes), it still is not
negligible.  We also derived masses from J and K or H and K
photometry, but we found them to be less consistent with H97, which is
the expected result if infrared excess is present.  Another possible
reason for the overestimated masses could be the different PMS tracks
used in H97 and this work.  H97 used the tracks by D'Antona \&
Mazzitelli (\cite{DanMaz94}), which are known to underestimate masses
(Hillenbrand \& White \cite{HillWhite2004}).  We prefer the tracks by
Baraffe et al.\ (\cite{Baraffe98}) and Siess et al.\
(\cite{Siess2000}) because they give magnitudes in infrared bands, so
no additional transformation is needed.  Hillenbrand \& White
(\cite{HillWhite2004}) show that the masses in the range 0.5 to
1.2$\,M_{\sun}$ predicted by the Baraffe et al.\ models are more
consistent with dynamical masses than those predicted by other models.
For higher masses, Hillenbrand \& White find good agreement between
predicted and dynamical masses for all models, including those of
Siess et al.

These mass estimates are good enough for our purposes, since they are
only used to classify the stars as intermediate- to high-mass
($M>2\,M_{\sun}$), low-mass ($0.1\,M_{\sun} < M < 2\,M_{\sun}$),
or sub-stellar candidates ($M < 0.1\,M_{\sun}$).
The number of systems that might be misclassified because of
inaccurate mass estimates is small and does not affect our
conclusions.

\subsection{Sensitivity Limits and Completeness}
\label{CompletenessSect}

\begin{figure}[t]
\centerline{\includegraphics[angle=270,width=\hsize]{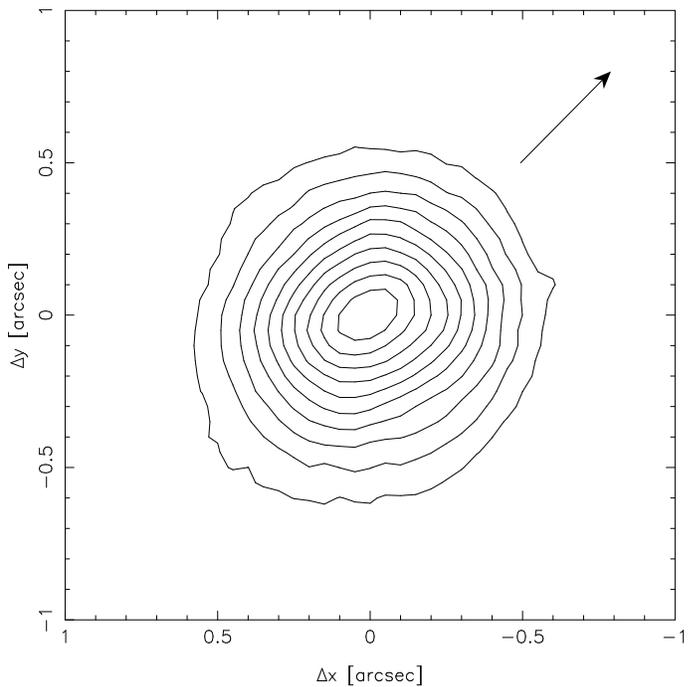}}
\caption{Our AO-corrected image of the star $42\farcs5$ south-east of
	\object{JW0950}, which was used as AO guide star.  At such
	large distances from the guide star, anisoplanatism leads to
	rather poor correction, which causes elongated star images
	with rather large FWHM.  The arrow indicates the direction
	to the guide star.}
\label{ElongExFig}
\end{figure}

The sensitivity of AO observations to close companions depends on many
factors, e.g.\ the telescope and instrument used, the atmospheric
conditions, the brightness of the guide star, and the distance from
the guide star to the target.  Since the images of target stars
separated by more than $\approx30''$ from the guide star are elongated
due to anisoplanatism (cf.\ Fig.~\ref{ElongExFig}), the sensitivity
for close companions depends also on the position angle.
All these factors vary considerably in our survey; it is therefore
impossible to give a single sensitivity limit.  Instead, we use a
statistical approach.

In Sect.~\ref{SingObsSect}, we describe how we measure the
completeness of the observation of one target star.  The result is a
limiting magnitude difference as function of separation and position
angle.  We convert this into a map giving the {\em probability} to
detect a companion as function of separation and magnitude difference.

In Sect.~\ref{ListObsSect}, we explain how the completeness maps of
several stars are combined into one map for the sample, and how the
corrections are computed that allow us to compare samples with
different completeness levels.


\begin{figure}[tp]
\centerline{\includegraphics[angle=90,width=1.1\hsize]{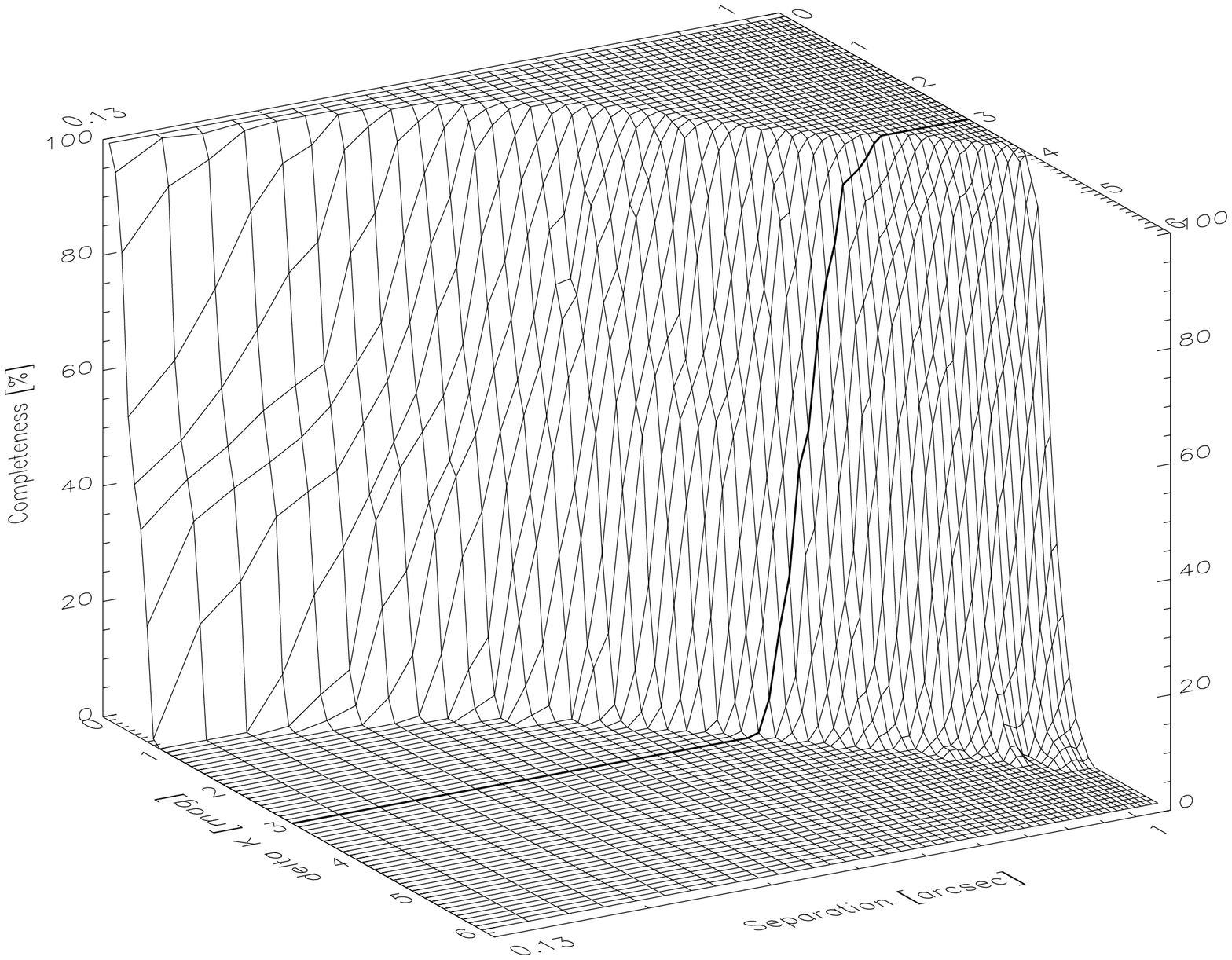}}
\centerline{\includegraphics[angle=270,width=\hsize]{4561f07b.ps}}
\caption{The completeness map of the observation of the star
	\object{JW0252}, presented as surface- (top) and contour-plot
	(bottom).  The thick line on the surface indicates a magnitude
	difference of $3^{\rm mag}$. The contour lines mark, from top
	to bottom, 100\,\%, 75\,\%, 50\,\%, 25\,\%, and 1\,\%
	completeness, the crosses show the limiting magnitudes that
	were found by visual inspection of images with artificial
	stars.  Similar maps were created for all targets in our
	sample, the completeness of the observation of JW0252 is on
	an average level for a low-mass star.}
\label{ComplExamplFig}
\end{figure}

\subsubsection{Completeness of a Single Observation}
\label{SingObsSect}

We start with a single target star and measure the sensitivity for
companions at various positions in the image.  In order to obtain
estimates for the sensitivity of all targets, an automated procedure
was used, which is based on the statistics of background
fluctuations.

For this method, we compute the standard deviation of the pixel values
in a $3\times8$ pixel box.  The short axis is oriented in radial
direction from the star, since the sensitivity varies over much
shorter scales in radial than in azimuthal direction.  Five sigma is
adopted as the maximum peak height of an undetected companion.  This
peak height is compared to the height of the central peak (i.e.\ that
of the primary star) to compute the magnitude difference.  The
procedure was repeated at 180 different position angles and 100
separations between 0\farcs02 and 2\arcsec.  This gives the limiting
magnitude difference as a function of separation and position angle.

We are not interested in the completeness limit at a particular
position angle.  Instead, we would like to know the probability to
detect a companion at a given separation and magnitude difference, for
example $0.7\arcsec$ and $\Delta K=1.5^{\rm mag}$, but at any position
angle.  We can safely assume that the distribution of companions in
position angle is uniform.  Then, the detection probability is
equivalent to the fraction of position angles where we can detect a
companion at $0.7\arcsec$ and $\Delta K=1.5^{\rm mag}$, i.e.\ the
fraction of position angles where the completeness limit at
$0.7\arcsec$ is fainter than $\Delta K=1.5^{\rm mag}$.

We repeat this counting of position angles for many different
separations and magnitude differences, which results in a map of the
probability to detect a companion as function of separation and
magnitude difference.  Figure~\ref{ComplExamplFig} shows the results
for the star JW0252 as an example.

\begin{figure*}[t]
\centerline{\includegraphics[angle=90,width=0.55\hsize]{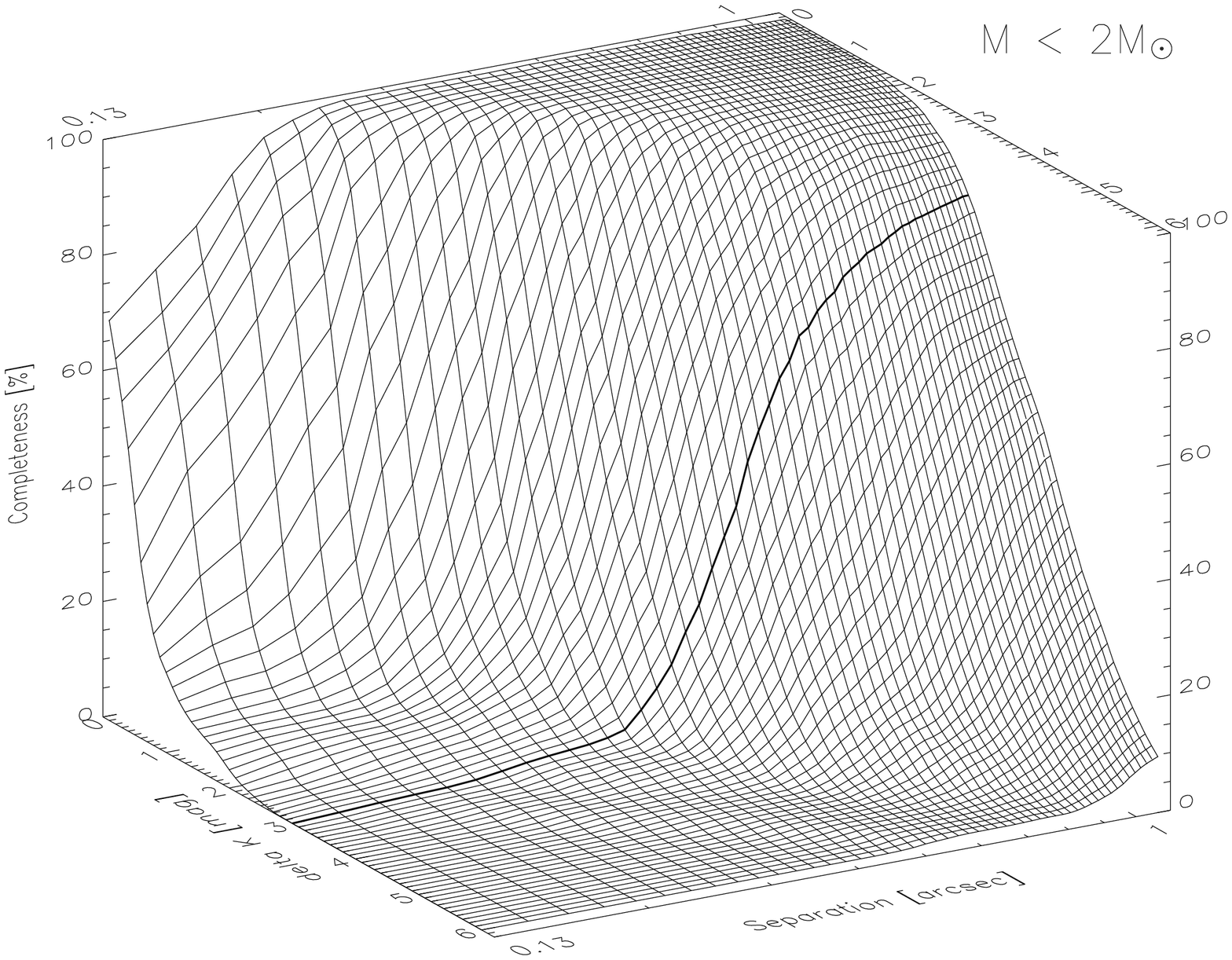}\hss
	    \includegraphics[angle=90,width=0.55\hsize]{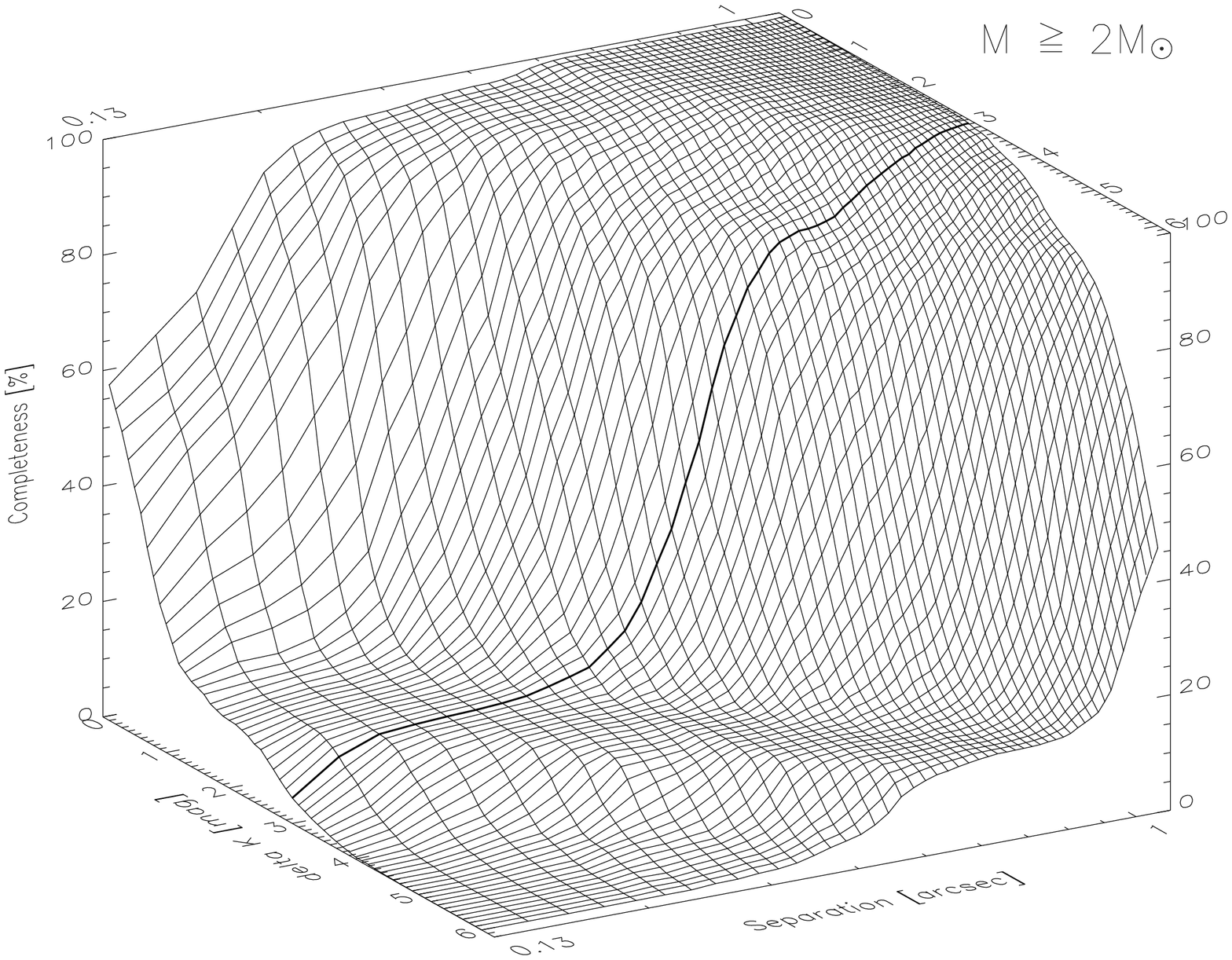}}
\centerline{\includegraphics[angle=270,width=0.47\hsize]{4561f08c.ps}\hfill
	    \includegraphics[angle=270,width=0.47\hsize]{4561f08d.ps}}
\caption{The completeness of the observations of low-mass (left) and
intermediate- to high-mass stars (right) in the
good sample, as function of separation and magnitude difference,
presented as a surface-plot (top) and a contour-plot (bottom).
On the surfaces, the thick line indicates a magnitude difference of
$3^{\rm mag}$.  It is interesting to note that the observations of
intermediate- to high-mass stars are generally more sensitive to faint
companions, but the observations of low-mass stars are slightly more
complete near the diffraction limit.}
\label{MbmMassFig}
\end{figure*}

To check whether these standard deviation calculations give good
approximation for the detection limits, tests with artificial stars
have been carried out.  We added artificial stars to our images (using
the target star PSF as template), and inspected the resulting images
in the same way as the original data.  The brightness of the
artificial stars was reduced to find the minimum magnitude difference
for a detectable companion.  Figure~\ref{ComplExamplFig} shows the
limits obtained in this way, they are comparable to the results of the
approach based on standard deviation calculations.  We therefore
conclude that, although tests with artificial stars might be a better
approach for determining the detection limits, the results based on
standard deviations agree quite well.  Computing standard deviations
at about 2 million positions is a matter of minutes, while visually
inspecting even a small number of images with artificial stars at
different positions is too time-consuming to be repeated for more than
200 stars.  Given the variations in the quality of the AO corrections
among our targets (mainly due to different magnitudes and separations
from the guide star), it is important to assess the completeness of
{\em all} observations and not just a few of them.

The final completeness maps are functions of two variables, separation
and magnitude difference.  Other binary surveys commonly specify the
completeness by giving the limiting magnitude difference as function
of separation.  This approach is not very useful for our survey since
the sensitivity varies considerably from primary to primary, therefore
we do not have a sharp limit in magnitude.  The transition zone
between 100\,\% and 0\,\% completeness is typically $1^{\rm mag}$ wide
(Fig.~\ref{ComplExamplFig}) for a single observation, and even wider
for the combined completeness of several stars (e.g.\
Fig.~\ref{MbmMassFig}).

We eliminated the dependence of the completeness on position angle,
but we pay a price: We no longer have a sharp completeness limit, but
a soft decline from 100\,\% to 0\,\% probability to detect a
companion.

\subsubsection{Completeness of several Observations}
\label{ListObsSect}

To combine the maps for the whole sample or selected subsamples into
an estimate for the completeness of the (sub)sample as function of
separation and magnitude difference, we simply average the
completeness.
To obtain the completeness of a subsample within a given range of
separation and $\Delta K$, we integrate the combined map within the
limits and divide the result by the area in separation-$\Delta K$
space.  This gives the completeness as a percentage between 0\,\% and
100\,\%.

In principle, we can extrapolate to 100\,\% completeness by dividing
the number of companions actually found by the completeness reached.
However, this method will only give accurate results if both the
distribution of separations and the distribution of magnitude
differences are flat.  We have no way to verify these assumptions, in
particular if we extrapolate to parts of the parameter space where we
are not able to detect companions.

In Sect.~\ref{CompOtherSect}, we compare our results to surveys of the
star-forming regions Taurus-Auriga and Scorpius-Centaurus.  These
star-forming regions are closer to the sun than the ONC, therefore
binaries with the same physical separation (in AU) have larger
projected separations (in arc-seconds) and are easier to detect.  When
we compare their results to our survey, the other survey is always
more sensitive for close and faint companions.  Instead of correcting
for incompleteness by adding companions we did not detect to the ONC
survey, we remove companions from the other surveys that would not
have been detected if they were in the ONC.  A prerequisite for this
method is, of course, that the authors give a complete list of
companions with separations and magnitudes or magnitude differences.

For comparisons of subsamples within the ONC we use a differential
method.  We construct a ``maximum completeness map'', which is at each
point in the separation-$\Delta K$ plane the maximum completeness of
the samples we wish to compare.  Then we extrapolate the number of
companions from the completeness of one sample to the maximum
completeness of both samples.  This difference is usually non-zero for
both samples, since the completeness of both samples is smaller than
the maximum completeness.  However, since the maximum completeness map
is only the envelope of the completeness maps of both subsamples,
the correction factors remain reasonably small, and the corrected
results do not contain large numbers of binaries that were in fact
never detected.  Fig.~\ref{MbmMassFig} shows the completeness maps of
low- and intermediate- to high-mass stars as an example.

\subsection{Chance Alignments}
\label{BgSect}

We observed our binary candidates only on one occasion;  there is no
way to tell from these data if two stars form indeed a gravitationally
bound system or if they are simply two unrelated stars that happen to
be close to each other projected onto the plane of the
sky.\footnote{Scally et al.\ (\cite{Scally99}) used proper motion data
for a statistical analysis of wide binaries in the ONC, but even this
additional information did not allow them to identify real binaries on
a star-by-star basis.}
Since the majority of the stars seen in the direction of the
  ONC are indeed cluster members, both stars in these pairs are
  usually in the cluster, chance alignments with background stars are
  relatively rare.  For this reason, we decided not to use
  the brightness of the companions to classify the binary candidates
  as physical pair or chance alignment.

{We can estimate how many of our binary candidates are
  chance alignments, with the help of the surface density of stars in
  our fields.  In the simplest case, the number of chance alignments
  we have to expect is the total area where we look for companions
  times the surface density of stars:
$$
	n_{\rm ch.a.} = N \cdot \pi r_{\rm out}^2 \cdot \Sigma,
$$
where $n_{\rm ch.a.}$ is the number of chance alignments, $N$ is the
number of primaries, $r_{\rm out}$ is the maximum radius where we
count a star as companion candidate, and $\Sigma$ is the surface
number density of stars on the sky.

However, we have to take into account the
sensitivity limits and (in)completeness derived in
Sect.~\ref{CompletenessSect}.  Since the completeness of our
observations is a function of radius and magnitude difference, we have
to integrate over both variables:
$$
  n_{\rm ch.a.} = \sum_{i=1}^N
	\int_0^{r_{\rm out}} \pi r \int_{m_i}^{m_i+\Delta m}
		p_i(r,m)\cdot\Sigma(m)\,{\rm d}m\,{\rm d}r,
$$
with $\Sigma(m)$ being the surface density of stars with magnitudes in
the range $m$ to $m+{\rm d}m$, and $p_i(r,m)$ the probability to
detect a star of magnitude $m$ at distance $r$ from target $i$.
Note that the integrals have to be computed for each target star
individually and the results added up, we cannot obtain the total by
multiplying with the number of targets $N$.

\begin{figure}[t]
\centerline{\includegraphics[angle=270,width=\hsize]{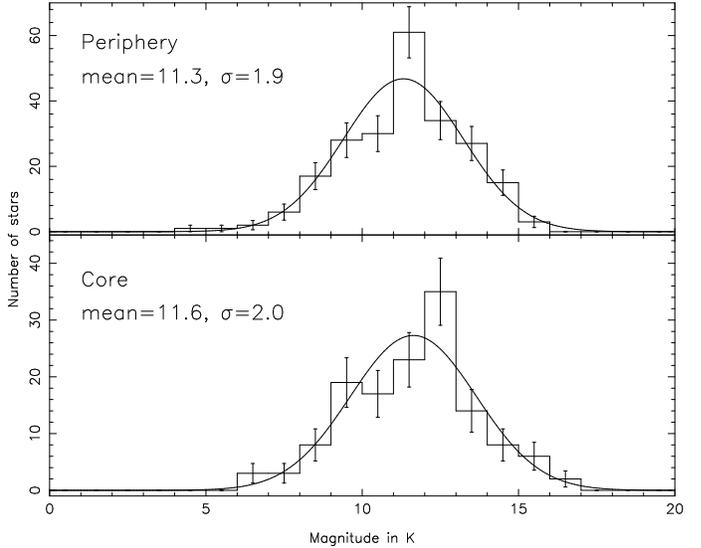}}
\caption{Number of stars as function of observed magnitude in the
K-band.  For the periphery, these are all stars within a radius of
$30\arcsec$ from one of the 52 guide stars.  The observations of the
core cover an area of about 2.2 square arcminutes.}
\label{KhistFig}
\end{figure}

\begin{table*}[t]
\caption[]{Comparison of low- and intermediate- to high-mass binaries,
	   corrections, and binary frequency.}
\label{csfmasstab}
\setlength{\tabcolsep}{4pt}
\begin{tabular}{lcc|cccr|cccr}
\noalign{\vskip1pt\hrule\vskip2pt}
Subsample    & Primary & Systems &
&  separations${}\le0.7\arcsec$\span\omit &   &	 & separations${}\le1.12\arcsec$\span\omit&\\
& Mass && Binaries & Chance~al. & Corr. & B.F. [\%] & Binaries & Chance~al. & Corr. & B.F. [\%]\\
\noalign{\vskip2pt\hrule\vskip2pt}
Good Sample  & $0.1\ldots 2M_{\sun}$ & 105 &  3 & 0.3 & $\times$1.32 & $ 3.4\pm 2.1$
					   &  5 & 0.9 & $\times$1.31 & $ 5.1\pm 2.7$\\
	     & $\ge2M_{\sun}$	     &  29 &  4 & 0.1 & $\times$1.01 & $13.5\pm 6.5$
					   &  4 & 0.4 & $\times$1.00 & $12.3\pm 6.4$\\
Large Sample & $0.1\ldots 2M_{\sun}$ & 228 &  9 & 1.2 & $\times$1.19 & $ 4.1\pm 1.6$
					   & 12 & 3.4 & $\times$1.24 & $ 4.7\pm 1.9$\\
	     & $\ge2M_{\sun}$	     &  48 &  8 & 0.3 & $\times$1.01 & $16.2\pm 5.4$
					   &  9 & 1.0 & $\times$1.00 & $16.7\pm 5.7$\\
\noalign{\vskip2pt\hrule\vskip2pt}
\end{tabular}
\end{table*}

To estimate the surface density as function of magnitude, we use the
targets detected in our own observations.  Because of the way the
sample was selected (cf.\ Sect.~\ref{ObsSect}), these are all the
sources in the 2MASS point-source catalog.  Our
high-spatial-resolution observations allow us to add a number of stars
that were either too faint for 2MASS, or in a region too crowded.  We
also removed a few spurious sources.  An additional advantage of using
our own observations to determine the stellar surface density is that
we do not have to apply corrections for incomplete detections of faint
sources.

One might argue that using the science targets for this purpose
introduces a bias because they are not randomly distributed in our
images, they are there because we chose to observe them.  However, we
did not decide to observe any particular target, we observed {\em all}
stars within 30\arcsec from a guide star (in the computation of the
surface density, we take into account that we effectively cover the
full field, not just the area of our AO-corrected images).

Figure~\ref{KhistFig} shows the number of stars in our sample as a
function of observed magnitude\footnote{Note that this is not the
  luminosity function of the cluster.  The conversion from observed to
  absolute magnitudes is not trivial, since it requires knowing the
  amount of extinction, which is not constant across the ONC.  We do
  not try to derive the luminosity function and we are not interested
  in it here, since the number of detected stars depends on the
  observed, not the absolute magnitude.}.
Because of the large difference in stellar surface density between
core and periphery of the cluster, we treat the two regions separately.
We approximate these distributions by Gau\ss\ functions with the
parameters given in the figure, the surface density is the number of
stars per magnitude interval divided by the area covered by the
observations.}

We then multiply the completeness maps derived for each star in the
last section by the number of stars per square arcsecond and magnitude
bin (in steps of 0.1$\rm^{mag}$).  The sum of the results is the
expected number of chance alignments with one target star, and the sum
of these expected numbers for a list of stars gives the number of
chance alignments we have to expect in total with this list of stars.

\subsubsection{Periphery}

Within circles of radius $1\farcs12$ around all 28 targets with masses
higher than 2$\,M_{\sun}$, we expect only $0.08\pm0.01$ chance
alignments.  Within the same separation from our 147 low-mass
targets, we expect $0.26\pm0.02$ chance alignments.  We expect
$0.006\pm0.001$ chance alignments with one of our 14 sub-stellar
candidates, which is in line with the fact that we find no
companion to these objects.

Inserting these numbers into the formula of the Binomial
distribution gives a probability of 92\,\% that we observe no chance
alignment with a intermediate- or high-mass star, and a probability of
77\,\% for no chance alignment with a low-mass star.  We can
therefore assume that all of our 4 intermediate- to high-mass and all
of the 7 low-mass binaries are indeed gravitationally-bound binaries.
This corresponds to a binary frequency of $(14\pm7)\%$ for the
intermediate- to high-mass and $(4.8\pm1.8)\%$ for the low-mass
systems.  Since there are no higher-order multiples, this is identical
to the multiplicity frequency (number of multiples divided by total
number of systems) and the companion star frequency (number of
companions divided by the total number of systems, CSF hereafter).


\subsubsection{Cluster Core}

In the cluster core, we carry out a similar analysis, but we have to
take into account the variations of the stellar surface density.
The surface density varies significantly from one target star to
  the next, we therefore cannot use the same $\Sigma$ for all
  targets. In principle, we have to derive $\Sigma(m)$ for each target
  individually.  However, the stellar surface density is not high
  enough to measure it as function of the stellar magnitude for each
  target field.  A region that contains enough stars to
  yield reasonable statistics would be too large to measure local
  density variations.  Therefore, we assume the shape of the function
  to be constant throughout the core region and determine it from all
  stars within the region (this is shown in the lower panel of
  Fig.~\ref{KhistFig}).  The result is scaled for each target to match
  the total number of stars found around it, which we obtained by
  counting the stars within 15\arcsec\ around each target.

Within radii of $1\farcs12$ from the 22 intermediate- and high-mass
stars in the core, we expect $1.0\pm0.3$ chance alignments.  For the
81 low-mass systems, the expected number is $3.1\pm1.0$, and for the
19 sub-stellar candidates, it is $0.19\pm0.05$.  The resulting
corrected binary frequencies are $(23\pm10)\%$ for the
intermediate- and high-mass stars, and $(3.6\pm3.2)\%$ for the
low-mass stars.



\section{Discussion}
\label{DiscuSect}

\subsection{Dependence of Binary Frequency on Primary Mass}
\label{MassSect}

First, we discuss the binary frequency of all stars as function of
primary mass, independent of their position in the cluster.  We study
two different subsamples based on the quality of the mass
determinations:

\begin{itemize}
\item The ``good sample'' consists of stars with masses given by H97
and a membership probability in the cluster of at least 50\,\%.  This
subsample contains 134 stars.

\item The ``large sample'' contains all stars with known masses,
either from H97 or from IR-photometry.  These masses are less
reliable, and we also include stars that are probably not cluster
members, but the statistical uncertainties are reduced by the larger
sample size of 275 systems.
\end{itemize}

Figure~\ref{MbmMassFig} shows the completeness-maps of low- and
intermediate- to high-mass stars in the good sample.
Table~\ref{csfmasstab} lists, for both samples and divided into
low- and intermediate- to high-mass stars, the number of binaries we
find, the number of chance alignments expected, the correction factor
to bring low- and intermediate- to high-mass stars to the same level
of completeness, and the resulting binary frequency.

The corrections for chance alignments are rather large if we count
binary candidates up to $1\farcs12$ separation -- one of the 5
low-mass binaries in the good sample, and 3 to 4 of the 12 low-mass
binaries in the large sample are probably chance alignments.  We
therefore decided to also study the statistics of closer binaries,
which are less affected by chance projections.  The upper separation
limit was chosen to be $0\farcs7$, which reduces the number of chance
alignments with low-mass stars in the large sample to about one.

We find the binary frequency of intermediate- to high-mass stars to be
always higher than that of low-mass stars, by a factor between 2.4 and
4.0.  The difference is statistically significant on the 1 to
2$\sigma$ level, where the ``good sample'' and the larger separation
limit have the lowest significance.  The good sample contains the
smaller number of systems, i.e.\ it has larger statistical
uncertainties.

The result that intermediate- and high-mass stars show a higher
multiplicity than low-mass stars was already found by previous studies
(e.g.\ Petr et al.\ \cite{Petr98}, Preibisch et al.\
\cite{Preibisch99}).  However, to our knowledge this is the first
study that shows a significant difference without applying large
corrections for undetected companions.  For example, Preibisch et al.\
(\cite{Preibisch99}) find 8 visual companions to 13 O- and B-stars.
Their result that the mean number of companions per primary star is at
least 1.5 is based on a correction factor of ${}\ga2.5$ for undetected
companions.  Our corrections for incompleteness raise the number of
low-mass companions (cf.\ Tab.~\ref{csfmasstab}), i.e.\ they reduce
the difference in the CSF between the two subsamples.

As a final remark in this section, we note that we find no binaries
among the sub-stellar candidates ($M<0.1\,M_{\sun}$) in our sample.
However, the completeness of these observations is limited in two
ways: We certainly did not find all sub-stellar objects in our fields,
and our sensitivity for companions is lower than for companions to
brighter objects.


\subsection{Comparison with other Regions}
\label{CompOtherSect}
\subsubsection{Main-sequence Stars}

In this section, we compare our findings to those of surveys of other
regions.  The binary survey most commonly used for comparison is the
work of Duquennoy \& Mayor (\cite{DM91}, hereafter DM91), who studied
a sample of 164 solar-type main-sequence stars in the solar
neighborhood.  They used spectroscopic observations, complemented by
direct imaging; therefore they give the distribution of {\em periods}
of the binaries.  Before we can compare this to our results, we have
to convert it into a distribution of projected separations.  We follow
the method described in K\"ohler (\cite{Koehler2001}).  In short, we
simulate 10 million artificial binaries with orbital elements
distributed according to DM91.  Then we compute the fraction of
those binaries that could have been detected by our observations,
i.e.\ those having projected separations between
$0\farcs13$ and $1\farcs12$ at the distance of the ONC (450\,pc).  The
result is that 21 binaries out of DM91's sample of 164 systems fall
into this separation range, which corresponds to a companion star
frequency of $(12.8\pm2.6)$\,\% (where the error is computed according
to binomial statistics).

DM91 report that their results are complete down to a mass ratio of
1:10.  The models of Baraffe et al.\ (\cite{Baraffe98}) show that for
low-mass stars on the 1\,Myr isochrone, this corresponds to a
magnitude difference of $3^{\rm mag}$ in the K-band.  Due to the
different method used (adaptive optics imaging vs.\ spectroscopy), our
survey is less complete than that of DM91.  With our data, we can find
only about 70\,\% of the binaries in the separation range
$0\farcs13\ldots1\farcs12$ and with magnitude differences up to
$3^{\rm mag}$ (Fig.~\ref{MbmMassFig}).  This means we would have found
a CSF of $(9.0\pm1.8)$\,\% for the DM91-sample.  In the ONC, the CSF
of low-mass stars in the good and the large sample are
$(3.9\pm2.1)$\,\% and $(3.8\pm1.5)$\,\%, respectively (after
subtracting chance alignments, but without corrections for
incompleteness, see Tab.~\ref{csfmasstab}).  This means the CSF in the
ONC is a factor of $2.3\pm1.0$ to $2.4\pm0.9$ lower than in the DM91
sample.


There are a number of M-dwarf surveys in the literature
 (Fischer \& Marcy \cite{FM92}, Leinert et al.\ \cite{Leinert97},
  Reid \& Gizis \cite{RG97}), none of them comprises a sample
 comparable in size to DM91.  Reid \& Gizis (\cite{RG97}, RG97)
 studied the largest sample so far, which contains 81 late K- or
 M-dwarfs.  Their list of binaries contains 4 companions in the
 separation range 60 -- 500\,AU.  They publish the separations and
 magnitudes of the binaries, which enables us to determine with the
 help of our completeness maps how many companions would have been
 detected if they were in Orion.  The result is about 2 companions,
 corresponding to a CSF of $(2.6\pm1.8)$\,\%, which is comparable to
 our result for stars in the ONC.  The lower CSF of M-dwarfs compared
 to solar-type stars reflects the deficit of M-dwarf binaries with
 larger separations (Marchal et al.\ \cite{Marchal03}).

It would be useful to compare our results for intermediate- to
high-mass stars in the ONC to the multiplicity of main-sequence stars
with similar masses.  However, we are not aware of a survey of
main-sequence B or A stars of similar quality and completeness as
DM91 for G stars.  A spectroscopic survey of B stars (Abt, Gomez, \&
Levy \cite{Abt90}) finds a binary frequency that is significantly
higher than among solar-type stars.  Mason et al.\ (\cite{Mason98})
find a binary frequency of 59 -- 75\,\% for O-stars in clusters,
and 35 -- 58\,\% for O-stars in the field.
They count binaries of {\em all} periods
and separations, using speckle observations combined with
spectroscopic and visual binaries from the literature.  This means we
cannot directly compare their numbers to our results for a limited
separation range.  Mason et al.\ apply no corrections for
incompleteness, therefore their numbers are rather lower limits.
Nevertheless, we can say that their results are in
qualitative agreement with our finding of a high binary frequency
among stars with masses${}\ga2\,M_{\sun}$ in clusters.

\subsubsection{Taurus-Auriga}

The star-forming region Taurus-Auriga is the best-studied
T~Association.  It is also the region where the overabundance of young
binary stars was discovered for the first time (Ghez et al.\
\cite{Ghez93}, Leinert et al.\ \cite{Leinert93}).  Here, we use the
binary surveys by Leinert et al.\ (\cite{Leinert93}) and K\"ohler \&
Leinert (\cite{Koehler98}), which together contain the largest
published sample of young stars in Taurus-Auriga.

The separation range 60 -- 500\,AU of our ONC-survey corresponds to
$0\farcs42$ -- $3\farcs6$ at the distance of Taurus-Auriga (140\,pc).
Within this range, the surveys by Leinert et al.\ and K\"ohler \&
Leinert are complete to a magnitude difference of about $6^{\rm mag}$
(cf.\ Fig.~2 in K\"ohler \& Leinert \cite{Koehler98}).  With our data,
we can find about 44\,\% of the binaries with magnitude differences up
to $6^{\rm mag}$.  Instead of multiplying the number of companions in
Taurus-Auriga by this percentage, we go back to the raw data of
the surveys in Taurus-Auriga and count the number of companions we
could have detected if they were in Orion, based on the completeness
estimated for our observations.  The result, after correction for
chance alignments and the bias induced by the X-ray selection
(K\"ohler \& Leinert \cite{Koehler98}), is 24 or 25 companions for the
completeness of our good and large sample, respectively.  With a
sample size of 174, this corresponds to a CSF of $(14\pm3)$\,\%.
In Orion, however, we find only a CSF of $(3.9\pm2.1)$\,\% and
$(3.8\pm1.5)$\,\%.  Thus, the CSF of low-mass stars
in Taurus-Auriga is higher by a factor of $3.5\pm1.6$ to $3.8\pm1.4$
compared to the ONC.

It is a well-known fact that the multiplicity in Taurus-Auriga is
about twice as high as among solar-type main-sequence stars.  With a
CSF in the ONC lower than among solar-type main-sequence stars by a
factor of about 2.4, we would expect the CSF in Taurus-Auriga to be
about 5 times higher than in the ONC.  The reason
why we find a smaller factor is the flux ratio distribution of the
binaries in Taurus-Auriga
(top panel of Fig.~8 in K\"ohler \& Leinert \cite{Koehler98}).  Many
of the companions would be too faint to be detected in our survey in
the ONC, therefore the overabundance of binaries in Taurus-Auriga is
less pronounced if we limit ourselves to the sensitivity of the
ONC-survey.

\subsubsection{Scorpius-Centaurus}

The Scorpius OB2 association is the closest and best-studied OB
association, in particular with respect to binaries.

Shatsky \& Tokovinin (\cite{Shatsky2002}) studied the multiplicity of
115 B-type stars in the Scorpius OB2 association, using coronographic
and non-coronographic imaging with the ADONIS system.  The separation
range of our survey corresponds to 0\farcs42 -- 3\farcs6 at the mean
distance of Scorpius OB2 (140\,pc).  Their list of physical companions
contains 12 objects in this separation range.  However, we would have
detected only about 4 of them if they were in the ONC, so the
detectable CSF is $(3.4\pm1.8)$\,\%.  We find $(12\pm6)$\,\% to
$(17\pm6)$\,\% binaries, which is higher with a statistical
significance of 1.3 to 2.3\,$\sigma$.

Kouwenhoven et al.\ (\cite{Kouwenhoven2005}) surveyed A star members
of Scorpius OB2 with ADONIS.  They adopt a distance of 130\,pc,
therefore the separation range of our survey corresponds to 0\farcs45
-- 3\farcs88.  Within these separations, they find 41 companions in
199 systems, of which 23 would have been detected by our observations.
This corresponds to a companion star frequency of $(11.5\pm2.3)$\,\%.
Our result of $(12\pm6)$\,\% to $(17\pm6)$\,\% is in agreement within
the error bars.

K\"ohler et al.\ (\cite{Koehler2000}) carried out a multiplicity
survey of T~Tauri stars in the Scorpius-Centaurus OB association.
Their targets are located in the Upper Scorpius part of the
association, which is at 145\,pc distance, therefore separations of
60 -- 500\,AU correspond to 0\farcs40 -- 3\farcs48.  They find 27
companions within this range in 104 systems, of which about 20 would
have been detected by us.  After correction for chance alignments,
this yields a CSF of $(18\pm4)$\,\%.  Comparing with our result in the
ONC of $(3.9\pm2.1)$\,\% to $(3.8\pm1.5)$\,\% leads to the conclusion
that the CSF of low-mass stars in Scorpius-Centaurus is higher than in
the ONC by a factor of about $5\pm2$.

There are a few studies of the multiplicity of very low-mass
objects in Upper Scorpius (Bouy et al.\ \cite{Bouy2006}, Kraus et al.\
\cite{Kraus2005}), but the limited completeness and sensitivity of our
observations of sub-stellar candidates does not allow a statistically
meaningful comparison.

\subsubsection{Summary of the Comparison with other Regions}

Table~\ref{sfrtab} gives an overview and summary of the results of
this section.

\begin{table}[t]
\caption[]{The CSF in several regions, reduced to the
completeness of our observations in the ONC (see text).}
\label{sfrtab}
\begin{tabular}{llc}
\noalign{\vskip1pt\hrule\vskip1pt}
\span\hfil Low-mass stars\span\\
\noalign{\vskip1pt\hrule\vskip1pt}
Region			& Reference	& CSF [\%]	\\
\hline
ONC (good sample)	& this work	& $3.9\pm2.1$	\\
ONC (large sample) 	& this work	& $3.8\pm1.5$	\\
Taurus-Auriga	   	& K\"ohler \& Leinert (\cite{Koehler98}) & $14\pm3$	\\
Scorpius-Centaurus 	& K\"ohler et al.\ (\cite{Koehler2000})  & $18\pm4$	\\
Main-sequence stars 	\\
\quad solar-type	& Duquennoy \& Mayor (\cite{DM91})	 & $9.0\pm1.8$	\\
\quad M-dwarfs		& Reid \& Gizis (\cite{RG97})		 & $2.6\pm1.8$	\\
\hline
\noalign{\vskip1pt\hrule\vskip2pt}
\span\hfil Intermediate- to High-mass stars\span\\
\noalign{\vskip1pt\hrule\vskip1pt}
Region		 	& Reference	& CSF [\%]	\\
\hline
ONC (good sample) 	& this work	& $12\pm6$	\\
ONC (large sample) 	& this work	& $17\pm6$	\\
Scorpius OB2		\\
\quad B-type stars & Shatsky \& Tokovinin (\cite{Shatsky2002})    &  $3.5\pm1.7$\\
\quad A-type stars & Kouwenhoven et al.\ (\cite{Kouwenhoven2005}) & $11.5\pm2.4$\\
\hline
\end{tabular}
\end{table}

{The data on low-mass stars show the result of the early surveys
(Leinert et al.\ \cite{Leinert93}, Ghez et al.\ \cite{Ghez93}), namely
the overabundance of binaries in Taurus-Auriga and Scorpius-Centaurus
compared to main-sequence stars.  Our results for binaries in the ONC
confirm that their frequency in the ONC is lower than among
solar-type main-sequence stars (e.g. Prosser et al.\ \cite{Prosser94},
Padgett et al.\ \cite{Padgett97}, Petr et al.\ \cite{Petr98}, Petr
\cite{PetrPhD}, Simon et al.\ \cite{Simon99}).  We find this
difference to be even more pronounced than in earlier studies.
However, we do not find a significant difference if we compare the
binary frequency of low-mass stars in the ONC and M-dwarfs in the
solar neighborhood (Reid \& Gizis\ \cite{RG97}).  This is not
surprising, since the median mass of our low-mass sample is
$0.3\,M_{\sun}$, most of the stars {\em are} M-dwarfs (cf.\ Fig.\
\ref{ColMagJHFig}).  This suggests that the difference in multiplicity
between Taurus and the ONC is not just a regional, but also a
selection effect: The surveys in Taurus-Auriga (Leinert et al.\
\cite{Leinert93}, Ghez et al.\ \cite{Ghez93}) and Scorpius-Centaurus
(K\"ohler et al.\ \cite{Koehler2000}) used Speckle interferometry,
which is limited to stars brighter than $\rm K\sim10^m$.  However,
selection effects cannot explain the difference between Taurus-Auriga
and solar-type main-sequence stars.

The data about intermediate- to high-mass stars present a somewhat
inconsistent picture: The results for the ONC is in good agreement
with those for A-type stars in the Scorpius OB2 association, which is
in line with the presumption that most of the intermediate- to
high-mass stars in our sample are
of spectral type A or later.  The binary frequency of B-type stars
appears to be lower, but this might be a selection effect.  These
numbers were obtained by taking into account the probability that the
companions could have been detected in our survey of the ONC.  This is
rather low for companions much fainter than the primary star, which
means it is harder to detect a companion to a B-star than detecting
an equally bright companion next to an A-star.  Shatsky \& Tokovinin
(\cite{Shatsky2002}) used a coronograph, which allowed them to find
fainter (and therefore more) companions than we could.  Their total
numbers (uncorrected for the sensitivity of our survey) agree quite
well with those for A-type stars (Kouwenhoven et al.\
\cite{Kouwenhoven2005}).
}

\begin{table*}
\caption[]{The numbers of binaries and systems, the corrections for
	chance alignments and incompleteness, and corrected binary
	frequency in the core and the periphery of the ONC.}
\label{csftab}
\setlength{\tabcolsep}{5pt}
\begin{tabular}{llcccccccccc}
\noalign{\vskip1pt\hrule\vskip2pt}
\span\span\span\span\span
binaries with separations${}\le1.12\arcsec$
\span\span\span\span\span\span\\
\noalign{\vskip1pt\hrule\vskip1pt}
Sample & Region & \span\span$0.1M_{\sun}\le M<2M_{\sun}$\span\span& \span\span$M\ge2M_{\sun}$\span\span\\
	&	& Bin. & Ch.~al. & Systems & Corr. & B.F. [\%]    & Bin. & Ch.~al. & Systems & Corr. & B.F. [\%]\\
\hline
Good Sample	& Core	    &   1 & 0.73 &  17 & $\times$1.02 & $ 1.6\pm 6.0$&  1 & 0.39 &  10 & $\times$1.14 & $ 6.9\pm10.9$\\
		& Periphery &   4 & 0.16 &  88 & $\times$1.33 & $ 5.8\pm 3.0$&  3 & 0.05 &  19 & $\times$1.02 & $15.8\pm 8.5$\\
\noalign{\smallskip}
Large Sample	& Core	    &   5 & 3.14 &  82 & $\times$1.02 & $ 2.3\pm 3.0$&  5 & 0.93 &  21 & $\times$1.09 & $21.1\pm10.2$\\
            	& Periphery &   7 & 0.26 & 146 & $\times$1.24 & $ 5.7\pm 2.2$&  4 & 0.08 &  27 & $\times$1.04 & $15.1\pm 7.1$\\
\noalign{\vskip1pt\hrule\vskip1pt}
\noalign{\vskip1pt\hrule\vskip2pt}
\span\span\span\span\span
binaries with separations${}\le0.7\arcsec$
\span\span\span\span\span\span\\
\noalign{\vskip1pt\hrule\vskip1pt}
Sample & Region & \span\span$0.1M_{\sun}\le M<2M_{\sun}$\span\span& \span\span$M\ge2M_{\sun}$\span\span\\
	&	& Bin. & Ch.~al. & Systems & Corr. & B.F. [\%]    & Bin. & Ch.~al. & Systems & Corr. & B.F. [\%]\\
\hline
Good Sample	& Core	    &   1 & 0.22 &  17 & $\times$1.03 & $ 4.7\pm 5.9$&  1 & 0.10 &  10 & $\times$1.22 & $11.1\pm11.6$\\
			& Periphery &   2 & 0.05 &  88 & $\times$1.31 & $ 2.9\pm 2.1$&  3 & 0.02 &  19 & $\times$1.03 & $16.2\pm 8.6$\\
\noalign{\smallskip}
Large Sample	& Core	    &   4 & 1.07 &  82 & $\times$1.02 & $ 3.7\pm 2.5$&  4 & 0.26 &  21 & $\times$1.14 & $20.3\pm 9.8$\\
             	& Periphery &   5 & 0.09 & 146 & $\times$1.30 & $ 4.4\pm 2.0$&  4 & 0.02 &  27 & $\times$1.06 & $15.6\pm 7.2$\\
\noalign{\vskip1pt\hrule}
\end{tabular}
\end{table*}

\begin{figure}[tp]
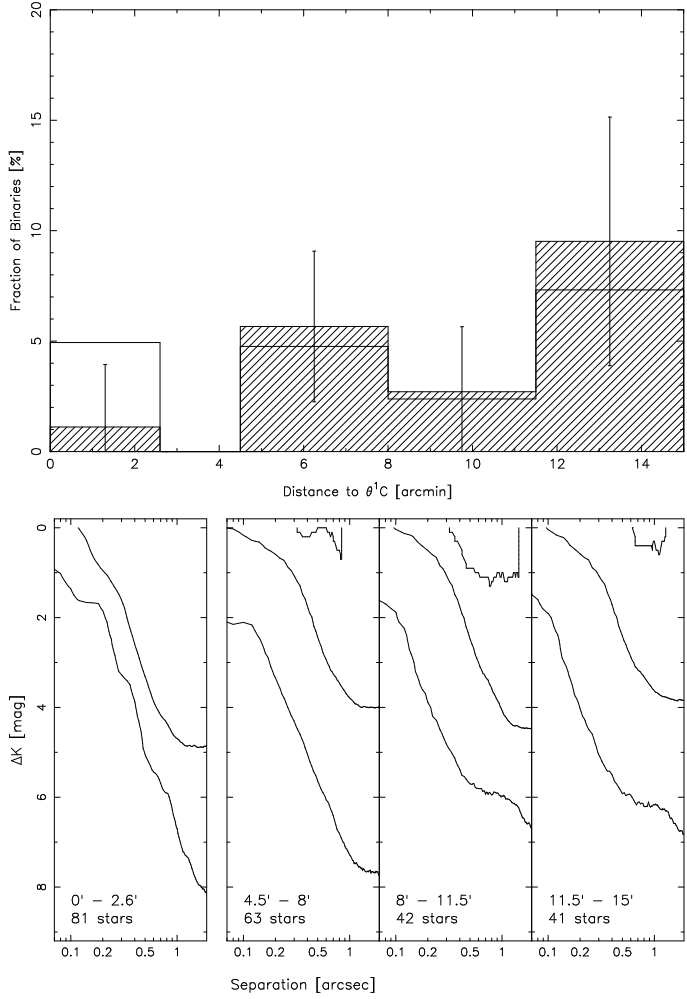

\centerline{\includegraphics[angle=270,width=\hsize]{4561f10a.ps}}
\medskip
\centerline{\includegraphics[angle=270,width=\hsize]{4561f10b.ps}}
\caption{Top: Companion star frequency vs.\ projected distance from
cluster center for all stars with $M < 2\,M_{\sun}$.  The outlined
histogram shows the uncorrected numbers, the hatched histogram and
the error bars show the results after correction for chance alignments
and relative incompleteness (see text).
Bottom: Completeness of the observations of these stars.
The contour lines mark 1\,\%, 50\,\%, and 100\,\% completeness.
Note that the sample in the cluster core never reaches 100\,\%
completeness at any separation, mainly because of targets close to the
edge of the frame.}
\label{CsfLomassFig}
\end{figure}
\begin{figure}[tp]
\centerline{\includegraphics[angle=270,width=\hsize]{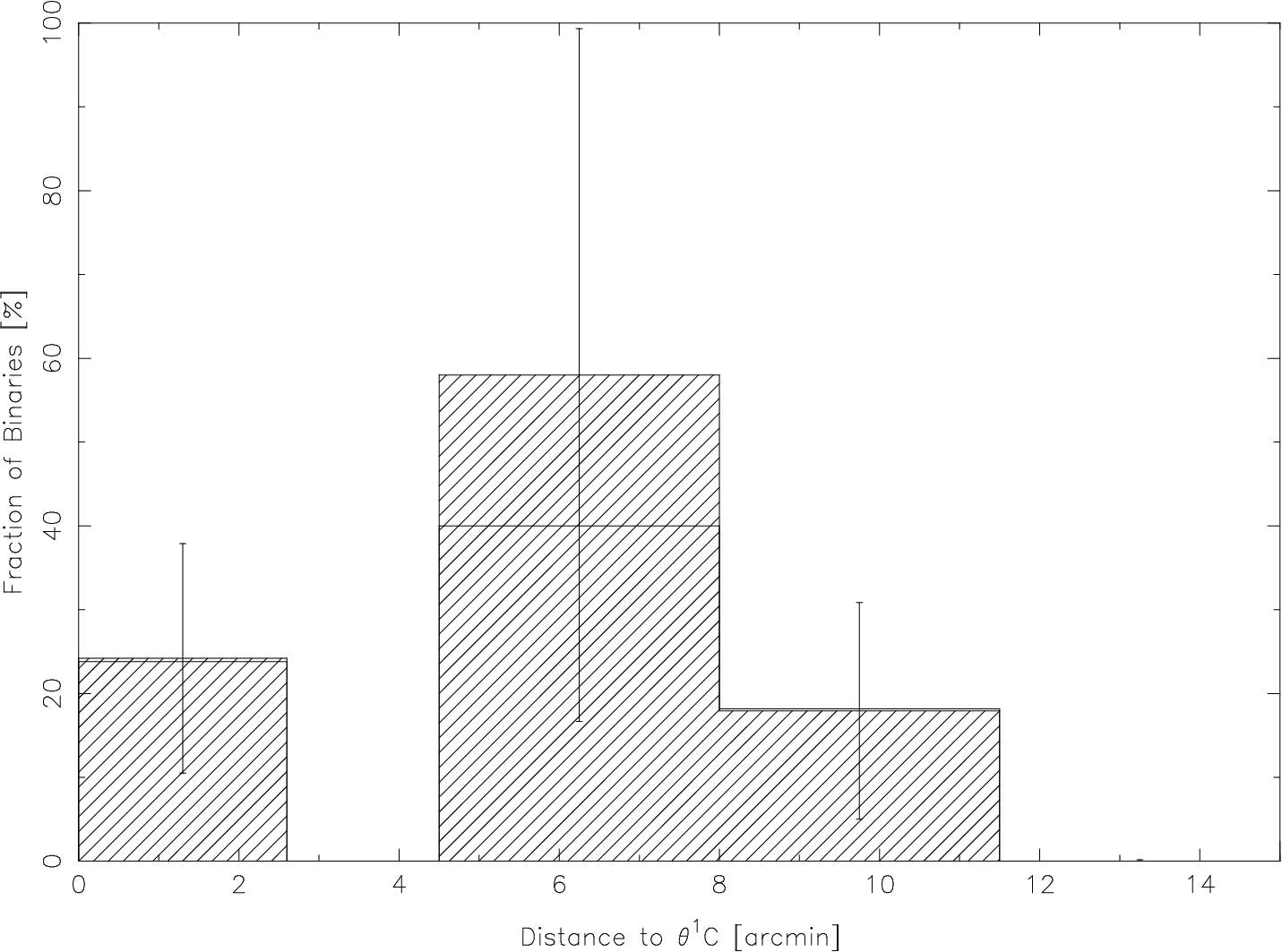}}
\medskip
\centerline{\includegraphics[angle=270,width=\hsize]{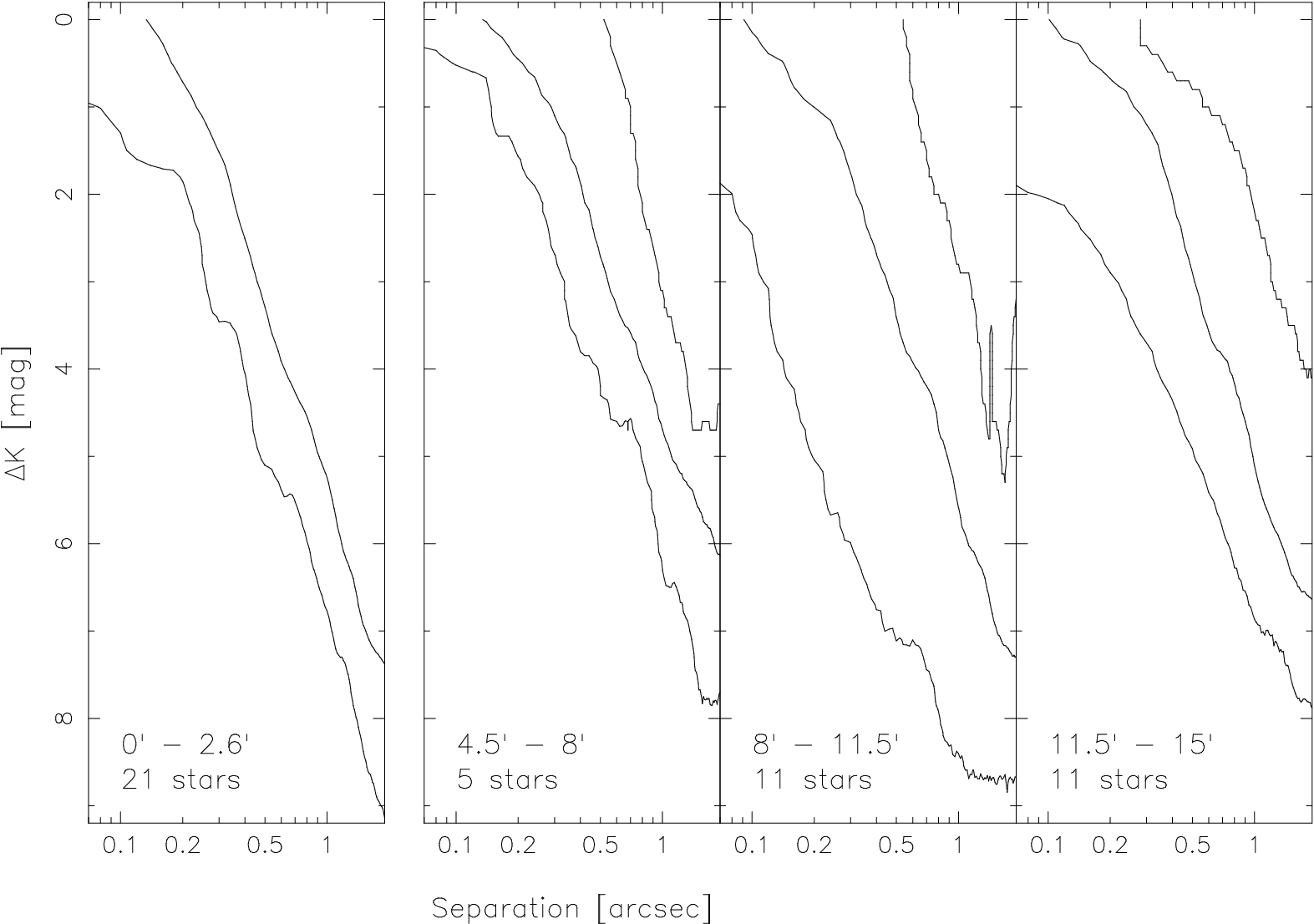}}
\caption{As Fig.~\ref{CsfLomassFig}, but for all stars with $M \ge
2\,M_{\sun}$.}
\label{CsfHimassFig}
\end{figure}

\subsection{Dependence of Binary Frequency on Distance from Cluster Center}

In this section, we study how the binary frequency depends on the
position of the star within the cluster.  This involves a number of
corrections to the raw numbers; the results are listed in
Tab.~\ref{csftab}, Fig.~\ref{CsfLomassFig}, and
Fig.~\ref{CsfHimassFig}.  As an example, we summarize how the numbers
for low-mass stars of the ``good sample'' in the cluster core were
derived.

All binaries and systems that match the criteria of the subsample
(i.e. mass between 0.1 and 2$\,M_{\sun}$, mass from H97, membership
probability $>50\,\%$, located less than $4\arcmin$ from
$\theta^1\rm\,C$) are counted.  Additionally, the probability for a
chance alignment is summed up, and the completeness of the whole
subsample is computed.  This completeness as a function of separation
and magnitude difference is compared to the completeness of the
low-mass stars of the good sample in the cluster periphery, in
order to derive the completeness correction factors.  Finally, we
subtract the number of chance alignments from the raw number of
companions, and multiply the result by the completeness correction to
obtain the binary frequency in table~\ref{csftab}.  Note that this is
not the binary frequency for 100\,\% completeness, but only for the
same completeness as the binary frequency of the corresponding
subsample of stars in the periphery.

As in Sect.~\ref{MassSect}, the corrections for chance alignments are
rather large, in particular in the cluster core.  Again, we study the
binary statistics for two upper separation limits, $0\farcs7$ and
$1\farcs12$.

The binary frequency of low-mass stars in the periphery is in most
cases somewhat higher than that of stars in the core
(Tab.~\ref{csftab} and Fig.~\ref{CsfLomassFig}).  The one exception
are binaries with separations${}<0\farcs7$ in the good sample, which
are more frequent in the core.  However, the number of binaries in
this subsample is too small to give any statistically meaningful
result.  The statistical significance of the other subsamples is also
quite small, at most $0.9\,\sigma$, where the largest subsample shows
the highest significance.  It is worth noting that this trend is not
present in the raw data, it appears only if the corrections for chance
alignments and incompleteness are applied.  We conclude that we find
only a small and statistically not very significant difference between
cluster core and periphery.

The results for stars with masses${}>2\,M_{\sun}$ are inconsistent
(Tab.~\ref{csftab} and Fig.~\ref{CsfHimassFig}).  Stars in the good
sample have higher multiplicity in the periphery, while the large
sample shows the opposite trend.  However, all the differences are
statistically insignificant, which leads to the conclusion that we
find no dependency of the binary frequency of intermediate- and
high-mass stars on the location in the cluster.  Given the small
numbers of binaries and systems in the subsamples, even this
conclusion should be taken with a grain of salt.

\subsection{ONC Periphery compared to Taurus-Auriga}

The result of a direct comparison between low-mass stars in the
periphery of the ONC and in Taurus-Auriga is much clearer: The CSF of
the stars in the periphery of the ONC after subtraction of chance
alignments is $(4.4\pm2.2)$\,\% (good sample) to $(4.6\pm1.8)$\,\%
(large sample).  In the separation range 60 -- 500\,AU, Leinert et
al.\ (\cite{Leinert93}) and K\"ohler \& Leinert (\cite{Koehler98})
find 39 companions in Taurus-Auriga, of which we would detect about 23
with the completeness of our observations.  This corresponds to a CSF
of $(13.4\pm2.6)$\,\%, and is higher than the CSF in the ONC by a
factor of $(3.1\pm1.3)$ and $(2.9\pm1.0)$.

\subsection{Implications for Binary Formation Theory}

Our results do not support the simple model by Kroupa et al.\
(\cite{Kroupa99}), which assumes that all stars in the ONC were born
in binary or multiple systems that were later destroyed by dynamical
interactions.  In this case, the binary frequency in the outer parts
of the cluster, where the dynamical timescales are too long for any
significant impact on the number of binaries since the formation of
the cluster, would be expected to be almost as high as in
Taurus-Auriga.  Our observations show that the binary frequency of
stars in the periphery is not significantly higher than in the cluster
core, and clearly lower than in Taurus.  Dynamical interactions may
explain the small difference we find between the binary frequencies in
the core and the periphery.  However, they are probably not the main
cause for the difference between the ONC and Taurus.

One way the rather homogeneous binary frequency throughout the
Orion Nebula Cluster can be explained by dynamical interactions is the
hypothesis that the cluster was much denser in the past.  This would
reduce the dynamical timescales and thus would not only disrupt more
binaries, but also lead to enough mixing to smear out differences in
binary frequency in different parts of the cluster.  However, while
there are indications that the ONC is expanding, it is hard to imagine
that it was so dense that the effect just described can explain the
observations.

Another possible explanation would be the hierarchical formation model
presented by Bonnell et al.~(\cite{BoBaVi2003}).  They used a
numerical simulation to follow the fragmentation of a turbulent
molecular cloud, and the subsequent formation and evolution of a
stellar cluster.  They show that the fragmentation of the molecular
cloud leads to the formation of many small subclusters, which later
merge to form the final cluster.  The number-density of stars in these
clusters is higher than in a monolithic formation scenario, which
results in closer and more frequent dynamical interactions.  The
relatively low binary frequency in the periphery of the ONC could
be explained if the binaries were already destroyed by
dynamical interactions in their parental subcluster.

However, the simplest explanation with our current knowledge of the
history of the ONC is still that {\em not\/} all stars there were born
in binary systems.  This means that the initial binary frequency in
the ONC was lower than, e.g., in Taurus-Auriga.  The reason for this
is probably related to the environmental conditions in the
star-forming regions' parental molecular clouds, like temperature or
the strength of the turbulence.  The observed anti-correlation between
stellar density and multiplicity does not imply that a high stellar
density causes a low binary frequency, but rather suggests that the
same conditions that lead to the formation of a dense cluster also
result in the formation of fewer binary and multiple stars.


\section{Summary and Conclusions}

We carried out a survey for binaries in the periphery of the Orion
Nebula Cluster, at 5 -- 15\,arcmin from the cluster center.  We
combine our data with the data of Petr (\cite{Petr98}) for stars in
the cluster core and find:
\begin{itemize}
\item The binary frequency of stars with $M > 2\,M_{\sun}$ is higher
  than that of stars with $0.1 < M < 2\,M_{\sun}$ by a factor of
  2.4 to 4.

\item The CSF of low-mass stars in the separation range 60 -- 500\,AU
  is comparable to M-dwarfs on the main sequence, but significantly
  lower than in other samples of low-mass stars: by a factor of 2 to 3
  compared to solar-type main-sequence stars, by a factor of 3 to 4
  compared to stars in the star-forming region Taurus-Auriga, and by a
  factor of about 5 compared to low-mass stars in Scorpius-Centaurus.
  The well-known overabundance of binaries in Taurus-Auriga is in this
  separation range largely caused by stars fainter than our detection
  limit, which have been excluded in the calculation of these factors.
  Therefore, the overabundance of binaries in Taurus-Auriga relative
  to main-sequence stars is not as pronounced in the factors relative
  to our results.

\item The CSF of intermediate- to high-mass stars in the ONC is
  significantly higher than that of B-type stars in Scorpius OB2,
  and comparable to that of A-type stars in the same region.
\end{itemize}

By comparing our results for stars in the periphery of the cluster
to those of Petr (\cite{Petr98}) for stars in the core we reach the
following conclusions:
\begin{itemize}
\item The binary frequency of low-mass stars in the periphery of the
  cluster is slightly higher than in the core, albeit with a low
  statistical significance of less than $1\,\sigma$.

\item The binary frequency of low-mass stars in the periphery of the
  ONC is lower than that of young stars in Taurus-Auriga, with a
  statistical significance on the $2\,\sigma$ level.

\item The binary frequency of stars with masses $>2\,M_{\sun}$ in
  the periphery is lower than in the center, but the difference is
  not statistically significant due to the small number of objects.
\end{itemize}

These results do not support the hypothesis that the initial binary
proportion in the ONC was as high as in Taurus-Auriga and was only
later reduced to the value observed today.  In that case, we would
expect a much higher number of binaries in the periphery than
observed.  There are models that can explain the observations with a
high initial binary frequency that was reduced by dynamical
interactions, e.g.\ a cluster that was much denser in the past, or a
hierarchical formation with many dense subclusters.  However, the
simplest explanation with our current knowledge is that the initial
binary frequency in the ONC was lower than in Taurus-Auriga.
This suggests that the binary formation rate is influenced by
environmental conditions, e.g.\ the temperature of the parental
molecular cloud.


\acknowledgements

The authors wish to recognize and acknowledge the very significant
cultural role and reverence that the summit of Mauna Kea has always
had within the indigenous Hawaiian community. We are most fortunate to
have the opportunity to conduct observations from this mountain.

This work has been supported in part by the National Science
Foundation Science and Technology Center for Adaptive Optics, managed
by the University of California at Santa Cruz under cooperative
agreement No.\ AST-9876783, and by the European Commission
Research Training Network ``The Formation and Evolution of Young
Stellar Clusters'' (HPRN-CT-2000-00155).

This publication makes use of data products from the Two Micron All
Sky Survey, which is a joint project of the University of
Massachusetts and the Infrared Processing and Analysis
Center/California Institute of Technology, funded by the National
Aeronautics and Space Administration and the National Science
Foundation.

The extensive report by an anonymous referee helped to improve the
paper considerably.

\begin{appendix}
\onecolumn
  \newpage
  \section{The Complete Target list}
  \setcounter{table}{5}%
\label{allstarstab}
\begin{longtable}{rlccccc}
\caption[]{Complete target list.}\\
\hline
No. & Name   & $\alpha_{2000}$ & $\delta_{2000}$ & H97 Mass & IR Mass & Guide star\\
\hline
\endfirsthead
\caption[]{Complete target list, continued.}\\
\hline
No. & Name   & $\alpha_{2000}$ & $\delta_{2000}$ & H97 Mass & IR Mass & Guide star\\
\hline
\endhead
\hline
\endfoot
1 & H97\,   5 & 5:34:29.2 & -5:23:57 & 1.73 & $1.4$ & JW0005\\
2 & 2MASS J0534292-052350 & 5:34:29.2 & -5:23:51 &      & $0.5$ & JW0005\\
3 & H97\,3082 & 5:34:28.9 & -5:23:48 &      & $0.1$ & JW0005\\
4 & H97\,   9 & 5:34:29.5 & -5:23:44 & 1.61 & $1.0$ & JW0005\\
5 & H97\,   8 & 5:34:29.4 & -5:23:38 & 0.10 & $0.1$ & JW0005\\
6 & H97\,3086 & 5:34:27.3 & -5:24:23 & 2.67 & $2.7$ & JW0005\\
7 & H97\,  14 & 5:34:30.2 & -5:27:27 & 2.08 & $1.7$ & JW0014\\
8 & H97\,3115 & 5:34:31.9 & -5:27:41 & 0.23 & $0.3$ & JW0014\\
9 & H97\,  27 & 5:34:33.9 & -5:28:25 & 1.41 & $1.5$ & JW0027\\
10 &  & 5:34:34.0 & -5:28:26 &      & $$ & JW0027\\
11 & 2MASS J0534336-052829 & 5:34:33.6 & -5:28:30 &      & $0.1$ & JW0027\\
12 & H97\,  18 & 5:34:31.6 & -5:28:28 & 0.29 & $0.3$ & JW0027\\
13 & H97\,  45 & 5:34:39.7 & -5:24:26 & 0.94 & $3.5$ & JW0045\\
14 & H97\,  40 & 5:34:38.2 & -5:24:24 & 0.21 & $0.2$ & JW0045\\
15 & 2MASS J0534382-052402 & 5:34:38.2 & -5:24:03 &      & $0.1$ & JW0045\\
16 & H97\,  54 & 5:34:41.5 & -5:23:57 & 0.14 & $0.1$ & JW0045\\
17 & H97\,  46 & 5:34:39.8 & -5:26:42 & 1.77 & $1.4$ & JW0046\\
18 & H97\,  51 & 5:34:40.7 & -5:26:39 & 0.36 & $0.3$ & JW0046\\
19 & H97\,  55 & 5:34:41.6 & -5:26:52 & 0.23 & $0.2$ & JW0046\\
20 & H97\,3113 & 5:34:39.5 & -5:27:17 &      & $0.6$ & JW0046\\
21 & H97\,  38 & 5:34:37.6 & -5:26:23 & 0.14 & $0.1$ & JW0046\\
22 & H97\,  50 & 5:34:40.8 & -5:22:43 & 0.34 & $3.0$ & JW0050\\
23 & H97\,  60 & 5:34:42.4 & -5:12:19 & 2.22 & $2.7$ & JW0060\\
24 & H97\,  59 & 5:34:42.3 & -5:12:38 & 0.16 & $0.1$ & JW0060\\
25 & H97\,  64 & 5:34:43.5 & -5:18:28 & 1.86 & $1.5$ & JW0064\\
26 &  & 5:34:43.4 & -5:18:28 &      & $$ & JW0064\\
27 &  & 5:34:45.4 & -5:18:54 &      & $$ & JW0064\\
28 & H97\,  75 & 5:34:45.1 & -5:25:04 & 1.50 & $2.0$ & JW0075\\
29 & H97\,  77 & 5:34:45.8 & -5:24:56 & 0.38 & $0.3$ & JW0075\\
30 & H97\,  81 & 5:34:46.3 & -5:24:32 & 0.32 & $0.6$ & JW0075\\
31 &  & 5:34:46.2 & -5:24:35 &      & $$ & JW0075\\
32 & H97\,  71 & 5:34:44.4 & -5:24:39 & 0.16 & $0.1$ & JW0075\\
33 & H97\,  62 & 5:34:42.8 & -5:25:17 & 1.38 & $0.6$ & JW0075\\
34 & H97\,  67 & 5:34:43.7 & -5:25:27 &      & $0.1$ & JW0075\\
35 & H97\, 108 & 5:34:49.9 & -5:18:45 & 3.59 & $3.5$ & JW0108\\
36 & H97\, 106 & 5:34:49.2 & -5:18:56 & 0.21 & $0.3$ & JW0108\\
37 & H97\,  98 & 5:34:48.7 & -5:19:08 & 0.18 & $0.2$ & JW0108\\
38 & H97\, 116 & 5:34:50.6 & -5:24:01 & 0.69 & $0.8$ & JW0116\\
39 &  & 5:34:50.7 & -5:24:03 &      & $$ & JW0116\\
40 & H97\, 114 & 5:34:50.4 & -5:23:36 & 0.41 & $     $ & JW0116\\
41 & H97\, 133 & 5:34:52.5 & -5:24:03 & 0.29 & $0.5$ & JW0116\\
42 & H97\, 125 & 5:34:51.9 & -5:24:18 & 0.20 & $0.2$ & JW0116\\
43 & H97\, 129 & 5:34:52.1 & -5:33:09 & 2.45 & $3.5$ & JW0129\\
44 &  & 5:34:53.2 & -5:33:09 &      & $$ & JW0129\\
45 & H97\, 115 & 5:34:50.3 & -5:32:55 & 0.11 & $0.1$ & JW0129\\
46 & 2MASS J0534502-053328 & 5:34:50.1 & -5:33:29 &      & $0.1$ & JW0129\\
47 & H97\, 153 & 5:34:55.2 & -5:30:22 & 3.08 & $3.0$ & JW0153\\
48 &  & 5:34:55.1 & -5:30:08 &      & $$ & JW0153\\
49 & H97\, 157 & 5:34:55.9 & -5:23:13 & 1.34 & $3.5$ & JW0157\\
50 &  & 5:34:54.5 & -5:23:02 &      & $$ & JW0157\\
51 & H97\, 171 & 5:34:57.0 & -5:23:00 & 0.17 & $     $ & JW0157\\
52 & H97\, 175 & 5:34:57.7 & -5:22:51 & 0.23 & $     $ & JW0157\\
53 & H97\, 163 & 5:34:56.7 & -5:11:33 & 0.68 & $0.8$ & JW0163\\
54 & H97\, 160 & 5:34:56.5 & -5:11:14 & 0.27 & $0.2$ & JW0163\\
55 &  & 5:34:56.5 & -5:11:04 &      & $$ & JW0163\\
56 & H97\, 165 & 5:34:56.4 & -5:31:36 & 2.30 & $1.5$ & JW0165\\
57 & H97\, 159 & 5:34:55.9 & -5:31:13 & 0.28 & $0.4$ & JW0165\\
58 & H97\, 221 & 5:35:02.3 & -5:15:48 & 1.52 & $>4.0 $ & JW0221\\
59 & H97\,3013 & 5:35:02.0 & -5:15:38 & 0.23 & $0.2$ & JW0221\\
60 & 2MASS J0535035-051600 & 5:35:03.5 & -5:16:00 &      & $0.1$ & JW0221\\
61 & H97\,5042 & 5:35:03.3 & -5:16:23 &      & $0.4$ & JW0221\\
62 & H97\, 232 & 5:35:02.9 & -5:30:01 & 1.45 & $5.0$ & JW0232\\
63 & H97\, 218 & 5:35:01.8 & -5:30:19 & 0.15 & $0.1$ & JW0232\\
64 & H97\, 235 & 5:35:03.4 & -5:29:26 &      & $0.5$ & JW0232\\
65 & H97\, 252 & 5:35:04.4 & -5:29:38 & 0.69 & $     $ & JW0232\\
66 & H97\, 256 & 5:35:04.5 & -5:29:36 & 0.21 & $     $ & JW0232\\
67 & H97\, 260 & 5:35:05.1 & -5:14:51 & 4.13 & $3.5$ & JW0260\\
68 & H97\, 282 & 5:35:06.0 & -5:14:25 & 0.12 & $0.3$ & JW0260\\
69 & H97\, 240 & 5:35:04.2 & -5:15:22 & 0.38 & $0.2$ & JW0260\\
70 & H97\, 364 & 5:35:11.5 & -5:16:58 & 3.04 & $2.5$ & JW0364\\
71 & H97\, 386 & 5:35:12.6 & -5:16:53 & 0.31 & $2.0$ & JW0364\\
72 & 2MASS J0535125-051633 & 5:35:12.4 & -5:16:34 &      & $3.0$ & JW0364\\
73 &  & 5:35:12.3 & -5:16:27 &      & $$ & JW0364\\
74 & H97\, 406 & 5:35:13.4 & -5:17:10 & 0.21 & $1.5$ & JW0364\\
75 & H97\, 408 & 5:35:13.4 & -5:17:17 & 0.23 & $0.2$ & JW0364\\
76 & 2MASS J0535131-051730 & 5:35:13.1 & -5:17:30 &      & $0.3$ & JW0364\\
77 & H97\, 407 & 5:35:13.4 & -5:17:31 & 0.57 & $0.5$ & JW0364\\
78 & H97\, 358 & 5:35:11.2 & -5:17:21 & 0.28 & $     $ & JW0364\\
79 & H97\, 367 & 5:35:11.8 & -5:17:26 &      & $0.3$ & JW0364\\
80 &  & 5:35:10.8 & -5:17:33 &      & $$ & JW0364\\
81 & H97\,5050 & 5:35:10.0 & -5:17:07 &      & $     $ & JW0364\\
82 & H97\,3166 & 5:35:09.2 & -5:16:57 & 0.29 & $     $ & JW0364\\
83 & H97\, 421 & 5:35:13.5 & -5:30:58 & 0.62 & $5.0$ & JW0421\\
84 & H97\, 416 & 5:35:13.4 & -5:30:48 & 0.21 & $0.2$ & JW0421\\
85 & H97\, 371 & 5:35:11.6 & -5:31:02 &      & $     $ & JW0421\\
86 & H97\, 428 & 5:35:13.7 & -5:30:24 & 0.25 & $0.3$ & JW0421\\
87 & H97\, 381 & 5:35:12.1 & -5:30:33 & 0.32 & $0.9$ & JW0421\\
88 & H97\, 585 & 5:35:18.4 & -5:16:38 & 0.35 & $2.0$ & JW0585\\
89 & H97\, 593 & 5:35:18.5 & -5:16:35 & 0.67 & $     $ & JW0585\\
90 & H97\, 574 & 5:35:18.1 & -5:16:34 & 0.15 & $     $ & JW0585\\
91 & H97\, 612 & 5:35:19.2 & -5:16:45 & 0.20 & $0.2$ & JW0585\\
92 & 2MASS J0535195-051703 & 5:35:19.5 & -5:17:03 &      & $3.0$ & JW0585\\
93 & H97\, 564 & 5:35:17.9 & -5:16:45 &      & $     $ & JW0585\\
94 & H97\, 525 & 5:35:16.8 & -5:17:03 & 0.17 & $     $ & JW0585\\
95 & H97\, 523 & 5:35:16.7 & -5:16:54 & 0.15 & $     $ & JW0585\\
96 & H97\,5047 & 5:35:17.4 & -5:16:57 &      & $     $ & JW0585\\
97 &  & 5:35:16.5 & -5:16:22 &      & $$ & JW0585\\
98 &  & 5:35:16.2 & -5:16:18 &      & $$ & JW0585\\
99 & H97\, 566 & 5:35:17.8 & -5:16:15 &      & $     $ & JW0585\\
100 &  & 5:35:17.4 & -5:16:14 &      & $$ & JW0585\\
101 &  & 5:35:17.6 & -5:16:17 &      & $$ & JW0585\\
102 & H97\, 601 & 5:35:18.7 & -5:16:15 & 1.17 & $     $ & JW0585\\
103 & H97\,5041 & 5:35:19.3 & -5:16:09 &      & $0.1$ & JW0585\\
104 & H97\,5040 & 5:35:19.2 & -5:16:10 &      & $     $ & JW0585\\
105 & H97\, 666 & 5:35:21.2 & -5:09:16 & 3.39 & $3.5$ & JW0666\\
106 & 2MASS J0535206-050902 & 5:35:20.5 & -5:09:03 &      & $0.2$ & JW0666\\
107 &  & 5:35:21.3 & -5:09:04 &      & $$ & JW0666\\
108 & H97\, 705 & 5:35:22.3 & -5:09:11 &      & $0.6$ & JW0666\\
109 & H97\, 671 & 5:35:21.3 & -5:09:42 & 0.29 & $0.1$ & JW0666\\
110 & H97\, 676 & 5:35:21.5 & -5:09:38 & 0.69 & $0.1$ & JW0666\\
111 &  & 5:35:21.7 & -5:09:45 &      & $$ & JW0666\\
112 & H97\, 677 & 5:35:21.5 & -5:09:49 & 0.22 & $0.2$ & JW0666\\
113 & H97\, 670 & 5:35:21.2 & -5:12:13 & 3.52 & $3.5$ & JW0670\\
114 &  & 5:35:20.1 & -5:12:11 &      & $$ & JW0670\\
115 & H97\, 747 & 5:35:23.7 & -5:30:47 & 1.75 & $1.4$ & JW0747\\
116 & H97\, 715 & 5:35:22.2 & -5:31:17 & 0.11 & $0.1$ & JW0747\\
117 & H97\, 788 & 5:35:25.5 & -5:30:38 & 0.17 & $0.3$ & JW0747\\
118 & H97\, 784 & 5:35:25.5 & -5:30:21 &      & $0.6$ & JW0747\\
119 & H97\, 787 & 5:35:25.3 & -5:30:22 & 0.30 & $     $ & JW0747\\
120 & H97\, 779 & 5:35:25.6 & -5:09:50 & 0.95 & $0.8$ & JW0779\\
121 & 2MASS J0535256-050942 & 5:35:25.5 & -5:09:42 &      & $0.1$ & JW0779\\
122 & H97\, 765 & 5:35:25.1 & -5:09:29 & 0.19 & $0.5$ & JW0779\\
123 & 2MASS J0535246-050926 & 5:35:24.5 & -5:09:27 &      & $0.0$ & JW0779\\
124 & 2MASS J0535268-050924 & 5:35:26.7 & -5:09:25 &      & $0.5$ & JW0779\\
125 & 2MASS J0535276-050937 & 5:35:27.5 & -5:09:37 &      & $>4.0 $ & JW0779\\
126 & H97\, 808 & 5:35:27.4 & -5:09:44 &      & $1.6$ & JW0779\\
127 & 2MASS J0535269-050954 & 5:35:26.9 & -5:09:54 &      & $0.1$ & JW0779\\
128 & 2MASS J0535269-051017 & 5:35:26.9 & -5:10:16 &      & $>4.0 $ & JW0779\\
129 & 2MASS J0535275-051008 & 5:35:27.5 & -5:10:08 &      & $0.1$ & JW0779\\
130 & 2MASS J0535251-051023 & 5:35:25.0 & -5:10:23 &      & $0.5$ & JW0779\\
131 & H97\, 790 & 5:35:26.2 & -5:08:40 & 2.04 & $1.5$ & JW0790\\
132 & H97\, 785 & 5:35:26.0 & -5:08:38 &      & $0.2$ & JW0790\\
133 & 2MASS J0535274-050903 & 5:35:27.3 & -5:09:03 &      & $0.1$ & JW0790\\
134 & 2MASS J0535250-050909 & 5:35:25.0 & -5:09:10 &      & $0.1$ & JW0790\\
135 & 2MASS J0535240-050906 & 5:35:24.0 & -5:09:07 &      & $0.1$ & JW0790\\
136 & H97\, 794 & 5:35:26.2 & -5:15:12 & 1.93 & $1.4$ & JW0794\\
137 &  & 5:35:26.4 & -5:15:06 &      & $$ & JW0794\\
138 & H97\, 805 & 5:35:27.0 & -5:15:45 &      & $0.3$ & JW0794\\
139 &  & 5:35:26.9 & -5:15:37 &      & $$ & JW0794\\
140 & H97\, 767 & 5:35:25.2 & -5:15:36 & 0.26 & $1.0$ & JW0794\\
141 & 2MASS J0535243-051457 & 5:35:24.3 & -5:14:59 &      & $0.1$ & JW0794\\
142 & H97\, 803 & 5:35:26.8 & -5:11:08 & 3.71 & $>4.0 $ & JW0803\\
143 & H97\, 822 & 5:35:28.1 & -5:11:37 & 0.27 & $0.4$ & JW0803\\
144 & H97\, 752 & 5:35:24.5 & -5:11:30 & 0.46 & $1.0$ & JW0803\\
145 & H97\, 771 & 5:35:25.3 & -5:10:49 & 1.15 & $0.6$ & JW0803\\
146 & H97\, 804 & 5:35:26.8 & -5:13:15 & 1.65 & $1.1$ & JW0804\\
147 & H97\, 818 & 5:35:27.4 & -5:35:20 & 0.32 & $0.9$ & JW0818\\
148 & H97\, 835 & 5:35:28.8 & -5:35:07 & 0.23 & $0.2$ & JW0818\\
149 & H97\, 866 & 5:35:31.2 & -5:15:33 & 1.95 & $6.0$ & JW0866\\
150 & H97\,3010 & 5:35:31.4 & -5:15:24 &      & $0.2$ & JW0866\\
151 & H97\,3011 & 5:35:31.5 & -5:15:24 &      & $     $ & JW0866\\
152 & H97\, 885 & 5:35:32.3 & -5:15:07 &      & $0.2$ & JW0866\\
153 & H97\, 867 & 5:35:31.2 & -5:18:56 & 0.30 & $1.0$ & JW0867\\
154 &  & 5:35:31.4 & -5:19:42 &      & $$ & JW0867\\
155 &  & 5:35:31.8 & -5:19:38 &      & $$ & JW0867\\
156 & H97\, 873 & 5:35:31.3 & -5:33:09 & 2.66 & $3.0$ & JW0873\\
157 &  & 5:35:31.2 & -5:33:11 &      & $$ & JW0873\\
158 & H97\, 847 & 5:35:29.7 & -5:32:54 & 1.09 & $1.6$ & JW0873\\
159 & 2MASS J0535325-053258 & 5:35:32.4 & -5:32:58 &      & $0.1$ & JW0873\\
160 & 2MASS J0535285-053304 & 5:35:28.5 & -5:33:06 &      & $0.1$ & JW0873\\
161 &  & 5:35:29.9 & -5:32:46 &      & $$ & JW0873\\
162 & H97\, 876 & 5:35:31.8 & -5:09:28 & 0.84 & $3.5$ & JW0876\\
163 & 2MASS J0535338-050905 & 5:35:33.9 & -5:09:04 &      & $0.1$ & JW0876\\
164 & H97\, 842 & 5:35:30.0 & -5:09:11 & 0.12 & $0.1$ & JW0876\\
165 & H97\, 887 & 5:35:32.3 & -5:31:11 & 2.97 & $3.0$ & JW0887\\
166 & H97\, 915 & 5:35:35.7 & -5:12:21 & 2.32 & $1.9$ & JW0915\\
167 &  & 5:35:35.5 & -5:12:16 &      & $$ & JW0915\\
168 & H97\, 928 & 5:35:37.3 & -5:26:40 &      & $1.1$ & JW0928\\
169 & 2MASS J0535381-052627 & 5:35:38.0 & -5:26:26 &      & $0.1$ & JW0928\\
170 & H97\, 912 & 5:35:35.1 & -5:26:54 & 0.23 & $0.3$ & JW0928\\
171 & H97\, 930 & 5:35:37.5 & -5:27:14 & 0.57 & $0.2$ & JW0928\\
172 & H97\, 950 & 5:35:40.4 & -5:27:02 &      & $0.6$ & JW0950\\
173 &  & 5:35:40.6 & -5:27:07 &      & $$ & JW0950\\
174 & H97\,5121 & 5:35:39.9 & -5:27:10 &      & $0.1$ & JW0950\\
175 & H97\, 962 & 5:35:42.4 & -5:27:33 & 0.58 & $0.7$ & JW0950\\
176 & H97\, 959 & 5:35:41.9 & -5:28:13 & 2.41 & $2.7$ & JW0959\\
177 &  & 5:35:42.0 & -5:28:11 &      & $$ & JW0959\\
178 & H97\, 954 & 5:35:41.2 & -5:27:51 & 0.23 & $0.3$ & JW0959\\
179 & H97\, 947 & 5:35:40.0 & -5:28:00 & 0.29 & $0.2$ & JW0959\\
180 & H97\, 944 & 5:35:39.5 & -5:27:51 & 0.18 & $0.2$ & JW0959\\
181 & H97\, 963 & 5:35:42.5 & -5:20:14 & 0.84 & $5.0$ & JW0963\\
182 & 2MASS J0535420-052005 & 5:35:41.5 & -5:20:06 &      & $0.1$ & JW0963\\
183 & 2MASS J0535427-051945 & 5:35:42.1 & -5:19:46 &      & $0.3$ & JW0963\\
184 & 2MASS J0535435-052047 & 5:35:43.0 & -5:20:47 &      & $0.4$ & JW0963\\
185 & 2MASS J0535436-052051 & 5:35:43.1 & -5:20:51 &      & $0.2$ & JW0963\\
186 & H97\, 967 & 5:35:42.9 & -5:13:46 & 1.85 & $1.5$ & JW0967\\
187 & H97\, 971 & 5:35:43.2 & -5:36:28 & 0.35 & $1.3$ & JW0971\\
188 & H97\, 974 & 5:35:44.1 & -5:36:40 & 0.14 & $0.3$ & JW0971\\
189 & H97\, 979 & 5:35:44.5 & -5:36:34 &      & $0.2$ & JW0971\\
190 & 2MASS J0535410-053622 & 5:35:41.0 & -5:36:25 &      & $0.8$ & JW0971\\
191 & H97\, 975 & 5:35:44.3 & -5:32:13 & 1.11 & $0.8$ & JW0975\\
192 & H97\, 973 & 5:35:43.3 & -5:32:09 & 0.12 & $0.1$ & JW0975\\
193 & H97\, 972 & 5:35:43.2 & -5:32:41 & 0.33 & $0.4$ & JW0975\\
194 & H97\, 992 & 5:35:46.9 & -5:17:57 &      & $     $ & JW0992\\
195 & H97\, 989 & 5:35:46.2 & -5:18:09 & 0.25 & $0.7$ & JW0992\\
196 & H97\, 983 & 5:35:45.5 & -5:18:14 &      & $0.5$ & JW0992\\
197 & H97\, 987 & 5:35:45.9 & -5:17:50 & 0.14 & $0.6$ & JW0992\\
198 & H97\, 993 & 5:35:47.1 & -5:17:44 & 0.14 & $     $ & JW0992\\
199 & 2MASS J0535485-051742 & 5:35:48.4 & -5:17:43 &      & $0.1$ & JW0992\\
200 & H97\, 997 & 5:35:47.4 & -5:16:58 & 1.83 & $1.4$ & JW0997\\
201 &  & 5:35:47.3 & -5:17:00 &      & $$ & JW0997\\
202 & H97\, 991 & 5:35:46.7 & -5:16:48 & 1.24 & $0.2$ & JW0997\\
203 &  & 5:35:47.1 & -5:16:44 &      & $$ & JW0997\\
204 & H97\,1015 & 5:35:50.4 & -5:28:35 & 2.52 & $>4.0 $ & JW1015\\
205 & H97\,1020 & 5:35:51.9 & -5:28:47 & 0.62 & $0.5$ & JW1015\\
206 & H97\,1041 & 5:35:58.0 & -5:12:55 & 3.92 & $3.5$ & JW1041\\
207 & H97\,1605 & 5:34:46.9 & -5:34:15 & 4.47 & $>4.0 $ & Par1605\\
208 & 2MASS J0534468-053423 & 5:34:46.8 & -5:34:24 &      & $0.2$ & Par1605\\
209 & H97\,  73 & 5:34:44.7 & -5:33:43 & 0.41 & $0.4$ & Par1605\\
210 & H97\,1744 & 5:35:05.0 & -5:12:16 & 4.81 & $3.5$ & Par1744\\
211 & 2MASS J0535056-051150 & 5:35:04.4 & -5:11:51 &      & $0.8$ & Par1744\\
212 & 2MASS J0535067-051145 & 5:35:05.5 & -5:11:46 &      & $0.3$ & Par1744\\
213 & H97\, 264 & 5:35:05.6 & -5:11:50 &      & $     $ & Par1744\\
214 & H97\,2074 & 5:35:31.3 & -5:16:03 & 16.3 & $>4.0 $ & Par2074\\
215 & H97\, 877 & 5:35:31.9 & -5:16:00 &      & $     $ & Par2074\\
216 & H97\, 882 & 5:35:32.1 & -5:16:04 &      & $     $ & Par2074\\
217 & H97\,3018 & 5:35:33.1 & -5:16:05 &      & $1.5$ & Par2074\\
218 & H97\, 892 & 5:35:32.8 & -5:16:05 & 0.65 & $1.1$ & Par2074\\
219 & H97\,3014 & 5:35:32.5 & -5:15:51 &      & $0.2$ & Par2074\\
220 & H97\,3019 & 5:35:29.8 & -5:16:07 &      & $0.4$ & Par2074\\
221 & H97\, 870 & 5:35:31.5 & -5:16:36 &      & $0.4$ & Par2074\\
222 & H97\, 875 & 5:35:31.7 & -5:16:39 &      & $0.2$ & Par2074\\
223 & H97\, 879 & 5:35:32.0 & -5:16:20 &      & $0.3$ & Par2074\\
224 & H97\,5043 & 5:35:32.3 & -5:16:27 &      & $0.1$ & Par2074\\
225 & H97\, 834 & 5:35:28.9 & -5:16:19 &      & $1.5$ & Par2074\\
226 & H97\, 837 & 5:35:29.5 & -5:16:34 & 0.34 & $0.6$ & Par2074\\
227 & H97\,2284 & 5:35:57.5 & -5:22:31 & 2.98 & $2.7$ & Par2284\\
228 &  & 5:35:57.3 & -5:22:30 &      & $$ & Par2284\\
\end{longtable}
\end{appendix}


\begin{thebibliography}{}

\bibitem[1990]{Abt90}
	Abt, H.~A., Gomez, A.~E., \& Levy, S.~G.
	1990, ApJSS 74, 551
\bibitem[1995]{AD95}
	Ali, B., \& Depoy, D.~L. 1995, AJ 109, 709

\bibitem[1998]{Baraffe98}
	Baraffe, I., Chabrier, G., Allard, F., Hauschildt, P.~H.
	1998, A\&A 337, 403
\bibitem[1998]{Bate98}
	Bate, M.~R., Clarke, C.~J., McCaughrean, M.~J.  1998,
	MNRAS 297, 1163
\bibitem[2003]{BeckSC03}
        Beck, T.~L., Simon, M., Close, L.~M.  2003,
	ApJ 583, 358
\bibitem[1998]{BonnellDavies98}
	Bonnell, I.~A., \& Davies, M.~B.\ 1998, MNRAS 295, 691
\bibitem[2003]{BoBaVi2003}
	Bonnell, I.~A., Bate, M.~R., Vine, S.~G.\ 2003, MNRAS 343, 413
\bibitem[1997]{Bouvier97}
	Bouvier, J., Rigaut, F., Nadeau, D.  1997,
	A\&A 323, 139
\bibitem[2006]{Bouy2006}
	Bouy, H., Mart{\'\i}n, E.~L., Brandner, W., Zapatero-Osorio,
	M.~R., B\'ejar, V.~J.~S., Schirmer, M., Hu\'elamo, N.,
	Ghez, A.~M.  2006, A\&A 451, 177 

\bibitem[1994]{DanMaz94}
	D'Antona, F., \& Mazzitelli, I.\ 1994, \apjs\ 90, 467
\bibitem[1999]{Duchene99a}
	Duch\^ene, G., 1999,
	A\&A 341, 547
\bibitem[1999]{Duchene99b}
	Duch\^ene, G., Bouvier, J., Simon, T.  1999,
	A\&A 343, 831
\bibitem[1991]{DM91}
	Duquennoy, A., \& Mayor, M. 1991, A\&A 248, 485 (DM91)
\bibitem[1994]{DurSterz94}
	Durisen, R.~H., \& Sterzik, M.~F.  1994, A\&A 286, 84

\bibitem[1992]{FM92}
        Fischer, D.~A., \& Marcy, G.~W. 1992, ApJ 396, 178

\bibitem[1993]{Ghez93}
	Ghez, A.~M., Neugebauer, G., Matthews, K. 1993,
	AJ 106, 2005
\bibitem[1997]{Ghez97}
	Ghez, A.~M., McCarthy, D.~W., Patience, J., Beck, T. 1997,
	AJ 481, 378

\bibitem[1997]{H97}
	Hillenbrand, L.~A. 1997, AJ\ 113, 1733, (H97)
\bibitem[1998]{Hillenbr98}
	Hillenbrand, L.~A., \& Hartmann, L.~W.  1998, ApJ\ 492, 540
\bibitem[2004]{HillWhite2004}
	Hillenbrand, L.~A., \& White, R.~J.  2004, ApJ\ 604, 741

\bibitem[1967]{Johnson67}
	Johnson, H.~L. 1967, ApJ 150, L39
\bibitem[1988]{JW88}
	Jones, B.~F., \& Walker, M.~F.  1988, AJ\ 95, 1755 (JW)

\bibitem[1998]{Koehler98}
	K\"ohler, R., \& Leinert, Ch.  1998, A\&A\ 331, 977
\bibitem[2000]{Koehler2000}
	K\"ohler, R., Kunkel, M., Leinert, Ch., \& Zinnecker, H.
	2000, A\&A 356, 541
\bibitem[2001]{Koehler2001}
	K\"ohler, R. 2001, AJ 122, 3325
\bibitem[2005]{Kouwenhoven2005}
	Kouwenhoven, M.~B.~N., Brown, A.~G.~A., Zinnecker, H.,
	Kaper, L., and Portegies Zwart, S.~F.
	2005, A\&A 430, 137
\bibitem[2005]{Kraus2005}
	Kraus, A.~L., White, R.~J., Hillenbrand, L.~A.\ 2005,
	ApJ 633, 452

\bibitem[1995]{Kroupa95}
	Kroupa, P.  1995,	MNRAS 277, 1491
\bibitem[1999]{Kroupa99}
	Kroupa, P., Petr, M., McCaughrean, M.  1999, New Astronomy 4, 495

\bibitem[1993]{Leinert93}
	Leinert, Ch., Zinnecker, H., Weitzel, N., et al.\
	1993, A\&A\ 278, 129
\bibitem[1997]{Leinert97}
	Leinert, Ch., Henry, T., Glindemann, A., McCarthy, D.~W.~, Jr.\
	1997, A\&A\ 325, 159
\bibitem[2003]{Liu2003}
	Liu, W.~M., Meyer, M.~R., Cotera, A.~S., Young,	E.~T.\
	2003, AJ\ 126, 1665
\bibitem[2005]{Luhman2005}
	Luhman, K.~L., McLeod, K.~K., Goldenson, N.\
	2005, ApJ\ 623, 1141

\bibitem[2003]{Marchal03}
	Marchal, L., Delfosse, X., Forveille, T., S\'egransan, D.,
	Beuzit, J.~L., Udry, S., Perrier, C., Mayor, M., Halbwachs,
	J.-L.\ 2003, In: ``Brown Dwarfs'', proceedings of IAU Symp.\
	No.\ 211, ed. E.~L.~Mart{\'\i}n, ASP Conference Series, p~311
\bibitem[1998]{Mason98}
	Mason, B.~D., Gies, D.~R., Hartkopf, W.~I., et al.\
	1998, AJ 115, 821
\bibitem[1994]{MJMStau94}
	McCaughrean, M.~J., \& Stauffer, J.~R.  1994, AJ\ 108, 1382
\bibitem[2001]{MJM2001}
	McCaughrean, M.~J.  2001,
	In: ``The Formation of Binary Stars'',
	proceedings of IAU Symp.\ No.\ 200, eds.\ H.~Zinnecker and
	R.~D.~Mathieu, ASP Conference Series, p.~169
\bibitem[2002]{Muench2002}
	Muench, A.~A., Lada, E.~A., Lada, Ch.~J., Alves, J.
	2002, \apj\ 573, 366

\bibitem[1997]{Padgett97}
	Padgett, D.~L., Strom, S.~E., \& Ghez, A.
	1997, \apj\ 477, 705
\bibitem[1954]{Parenago}
	Parenago, P.~P. 1954, Trudy Sternberg Astron.\ Inst., 25
\bibitem[1998]{Patience98}
	Patience, J., Ghez, A.~M., Reid, I.~N., Weinberger, A.~J.,
	Matthews, K.  1998, AJ 115, 1972
\bibitem[1998]{Petr98}
	Petr, M.~G., Coud\'e du Foresto, V., Beckwith, S.~V.~W.,
	Richichi, A., McCaughrean, M.~J.  1998, ApJ 500, 825.
\bibitem[1998]{PetrPhD}
	Petr, M.~G. 1998, PhD Thesis, University of Heidelberg.
\bibitem[1999]{Preibisch99}
	Preibisch, Th., Balega, Y., Hofmann, K.-H., Weigelt, G.,
	Zin\-necker, H.  1999, New Astronomy 4, 531
\bibitem[2001]{Preibisch2001}
	Preibisch, Th., Weigelt, G., Zin\-necker, H.  2001,
	In: ``The Formation of Binary Stars'',
	proceedings of IAU Symp.\ No.\ 200, eds.\ H.~Zinnecker and
	R.~D.~Mathieu, ASP Conference Series, p.~69
\bibitem[1994]{Prosser94}
	Prosser, C.~F., Stauffer, J.~R., Hartmann, L., et al.\
	1994, ApJ 421, 517.

\bibitem[1997]{RG97}
	Reid, I.~N., \& Gizis, J.~E.\ 1997, AJ\ 113, 2246

\bibitem[1999]{Scally99}
	Scally, A., Clarke, C., McCaughrean M.~J.  1999,
	MNRAS 306, 253.
\bibitem[2003]{Schertl03}
	Schertl, D., Balega, Y.~Y., Preibisch, Th., Weigelt, G.
	2003, A\&A 402, 267
\bibitem[2002]{Shatsky2002}
	Shatsky, N., \& Tokovinin, A.\ 2002, A\&A 382, 92.
\bibitem[2000]{Siess2000}
	Siess, L., Dufour, E., Forestini, M.  2000,
	\aap\ 358, 593
\bibitem[1999]{Simon99}
	Simon, M., Close, L.~M., Beck, T.~L. 1999,
	AJ 117, 1375.
\bibitem[2003]{Sterzik2003}
	Sterzik, M.~F., Durisen, R.~H., Zinnecker, H.
	2003, A\&A 411, 91

\bibitem[1997]{Wielen97}
	Wielen, R.  1997, A\&A  325, 367

\end{thebibliography}
\end{document}